\documentclass[11pt,draftclsnofoot,journal,onecolumn]{IEEEtran}

\IEEEoverridecommandlockouts

\usepackage[dvips]{graphicx}
\usepackage{subfigure}
\usepackage{amsmath}
\usepackage{amssymb}
\usepackage{url}
\usepackage{cite}
\usepackage{cases}
\usepackage{algorithmic}
\usepackage{setspace}
\usepackage[usenames]{color}
\usepackage{flushend}
\usepackage[ruled]{algorithm2e}

\newtheorem{theorem}{Theorem}
\newtheorem{corollary}{Corollary}
\renewcommand{\IEEEQED}{\IEEEQEDopen}

\newcounter{MYtempeqncnt}

\begin{document}

\title{Primary User Traffic Estimation for Dynamic Spectrum Access}
\author{Wesam Gabran, Chun-Hao Liu, Przemys{\l}aw Pawe{\l}czak, and Danijela Cabric
\thanks{Copyright\copyright~2012 IEEE. Personal use of this material is permitted. However, permission to use this material for any other purposes must be obtained from the IEEE by sending a request to pubs-permissions@ieee.org.}
\thanks{Wesam Gabran was with with the Department of Electrical Engineering, University of California, Los Angeles. He is currently with Broadcom Corporation, 5300 California Ave, Irvine, CA 92617, USA (email: wgabran@broadcom.com).}
\thanks{Chun-Hao Liu and Danijela Cabric are with the Department of Electrical Engineering, University of California, Los Angeles, 56-125B Engineering IV Building, Los Angeles, CA 90095-1594, USA (email: \{liuch37, danijela\}@ee.ucla.edu).}
\thanks{Przemys{\l}aw Pawe{\l}czak is with the Fraunhofer Institute for Telecommunications, Heinrich Hertz Institute, Einsteinufer 37, 10587 Berlin, Germany (email: przemyslaw.pawelczak@hhi.fraunhofer.de).}
\thanks{This work has been supported by the National Science Foundation under CNS grant 1117600 and the German Federal Ministry of Economics and Technology under grant 01ME11024.}}

\maketitle

\begin{abstract}
Accurate estimation of licensed channel Primary User's (PU) temporal statistics is important for Dynamic Spectrum Access (DSA) systems. With accurate estimation of the mean duty cycle, $u$, and the mean off- and on-times of PUs, DSA systems can more efficiently assign PU resources to its subscribers, thus, increasing channel utilization. This paper presents a mathematical analysis of the accuracy of estimating $u$, as well as the PU mean off- and on-times, where the estimation accuracy is expressed in terms of the Cram\'{e}r-Rao bound on the mean squared estimation error. The analysis applies for the traffic model assuming exponentially distributed PU off- and on-times, which is a common model in traffic literature. The estimation accuracy is quantified as a function of the number of samples and observation window length, hence, this work provides guidelines on traffic parameters estimation for both energy-constrained and delay-constrained applications. For estimating $u$, we derive the mean squared estimation error for uniform, non-uniform, and weighted sample stream averaging, as well as maximum likelihood (ML) estimation. The estimation accuracy of the mean PU off- and on-times is studied when ML estimation is employed. Besides, the impact of spectrum sensing errors on the estimation accuracy is studied analytically for the averaging estimators, while simulation results are used for the ML estimators. Furthermore, we develop algorithms for the blind estimation of the traffic parameters based on the derived theoretical estimation accuracy expressions. We show that the estimation error for all traffic parameters is lower bounded for a fixed observation window length due to the correlation between the traffic samples. On the other hand, the impact of spectrum sensing errors on the estimation error of $u$ can be eliminated by increasing the number of traffic samples for a fixed observation window length. Finally, we prove that for estimating $u$ under perfect knowledge of either the mean PU off- or on-time, ML estimation can yield the same estimation error as weighted sample averaging using only half the observation window length.
\end{abstract}

\section{Introduction}
\label{sec:Introduction}

Spectrum sensing is the cornerstone of Dynamic Spectrum Access (DSA)~\cite{Zhao_sigprocmag_2007} where Secondary Users (SUs) search for, and operate on, licensed spectrum that is temporarily vacant. The SUs have to sense for the presence of Primary (licensed) Users (PUs) on the targeted spectral bands before utilizing these radio resources. The PU channel utilization patterns are stochastic in nature~\cite{wellens_phycom_2009}. Consequently, acquiring knowledge about the PU traffic statistics can improve the performance of SU channel selection algorithms, for example~\cite{rashid_twc_2009}, and help in achieving more efficient resource allocation, for example~\cite{ngo_tvt_2010}, in DSA systems.
\IEEEpubidadjcol

\subsection{The Need for Accurate PU Traffic Estimation: an Example}
\label{sec:motovation}

\begin{figure}
\centering
\subfigure[$u_{\forall i}=0.8$]{\includegraphics[width=0.49\columnwidth]{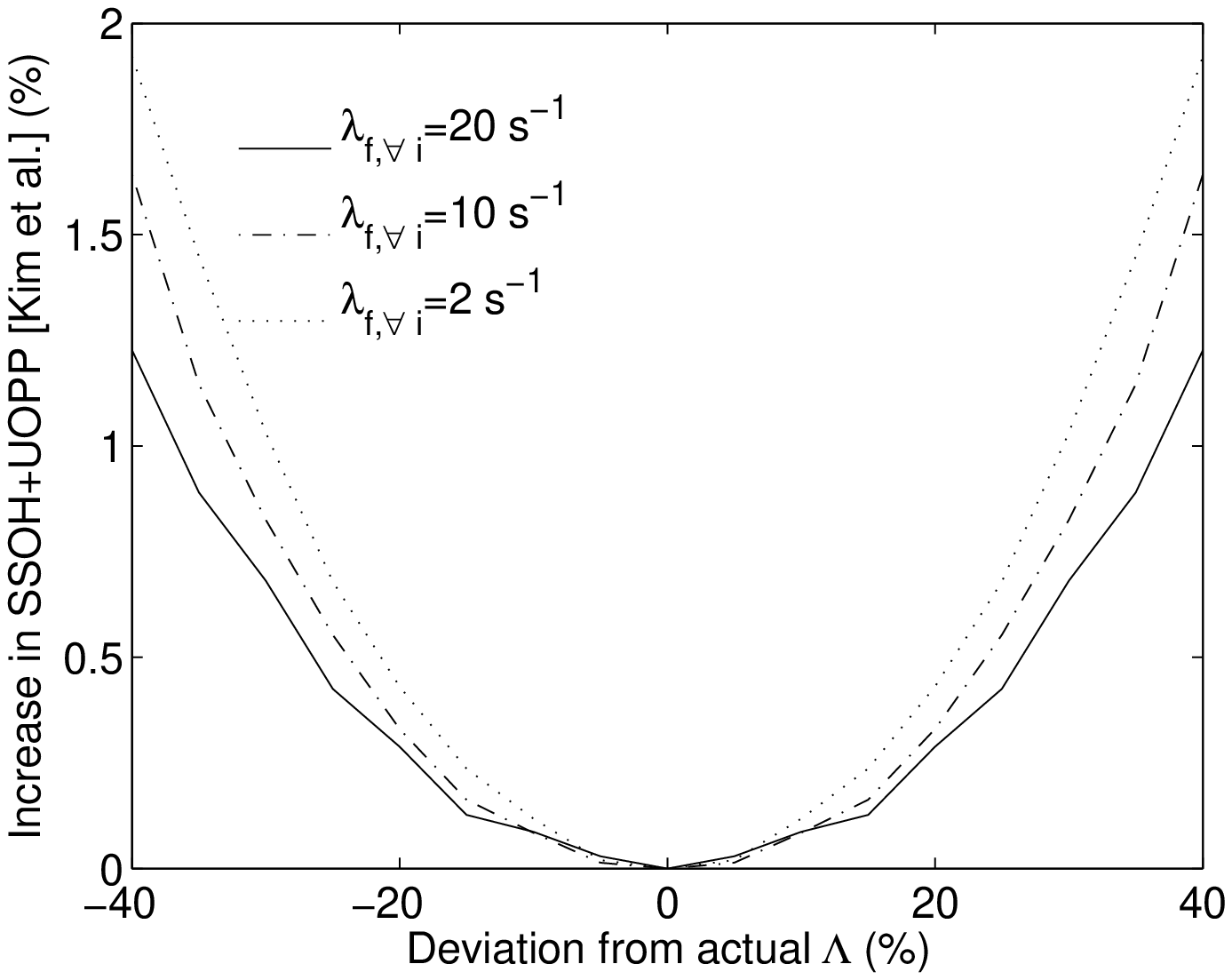}\label{fig:mac_lambda}}
\subfigure[$\lambda_{f,\forall i}=10$\,s$^{-1}$]{\includegraphics[width=0.49\columnwidth]{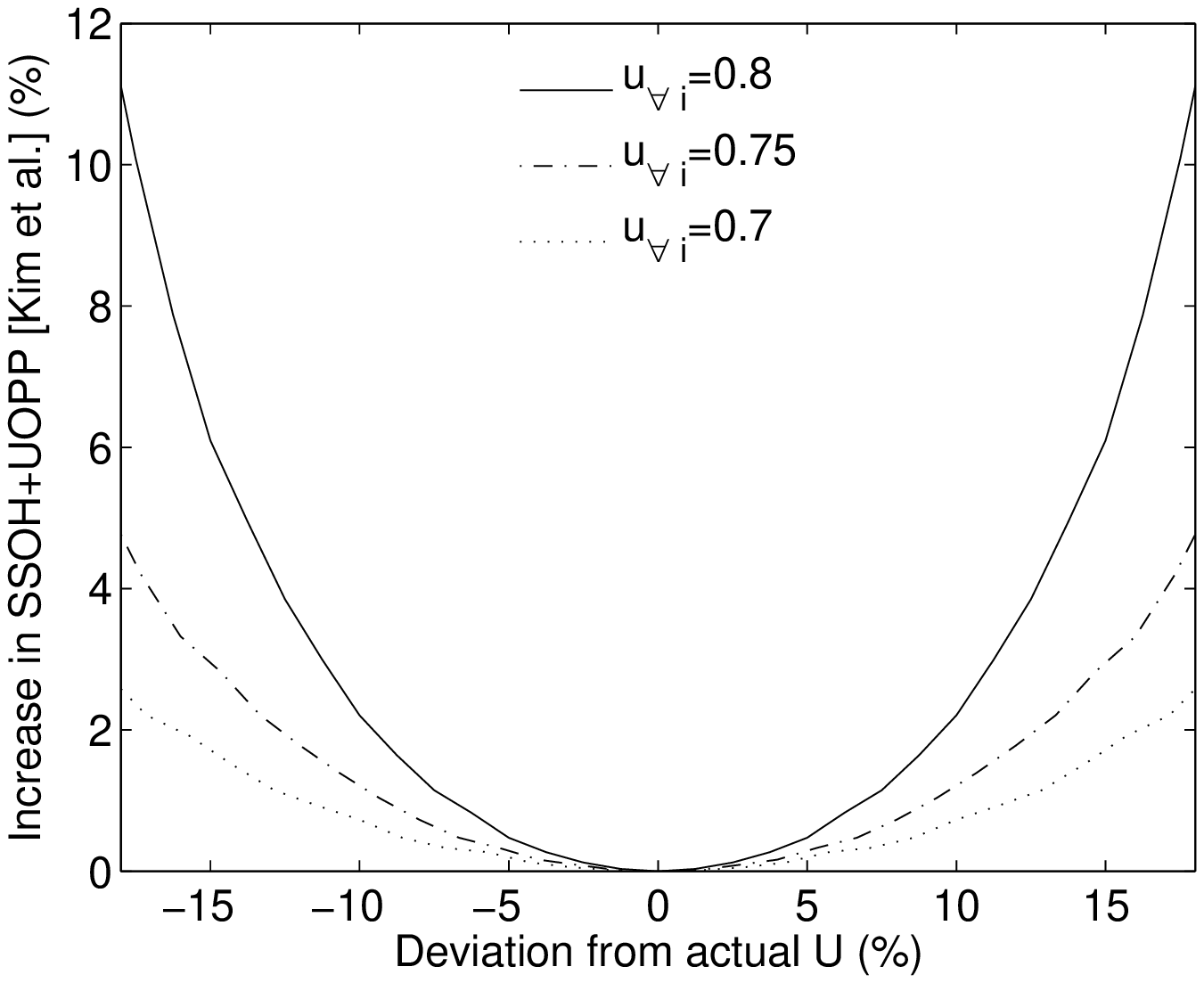}\label{fig:mac_u}}
\caption{Impact of inaccurate estimation of PU departure rate for channel $i$, $\lambda_{f,i}$, (Fig.~\ref{fig:mac_lambda}) and mean PU duty cycle for channel $i$, $u_{i}$, (Fig.~\ref{fig:mac_u}) on the performance of the multi-channel MAC protocol proposed by Kim et al. in~\cite{kim_tmc_2008}. The results were obtained as follows. With the number of PU channels, $N$, set to 2 and sensing time set to 1.6\,ms, vector $\Lambda=\{\lambda_{f,1},\cdots,\lambda_{f,N}\}$ in Fig.~\ref{fig:mac_lambda}, ($U=\{u_{1},\cdots,u_{N}\}$ in Fig.~\ref{fig:mac_u}), was shifted from the actual value (by the same factor) positively for the first channel and negatively for the second channel. Then, the corresponding optimal sensing period was calculated for the actual and erroneous $\Lambda$ and $U$ using~\cite[Eq. (1)]{kim_tmc_2008}. The resulting percentage increase in SSOH+UOPP was calculated for the case of having erroneous $\Lambda$, (Fig.~\ref{fig:mac_lambda}), and erroneous $U$, (Fig.~\ref{fig:mac_u}). For both figures inter-sampling granularity and maximum inter-sampling period were set to 0.5\,ms and 0.1\,s, respectively.}
\label{fig:mac_example}
\end{figure}

The multi-channel Medium Access Control (MAC) protocol proposed in~\cite{kim_tmc_2008} is a good example for showing the importance of PU traffic parameters estimation. In the proposed MAC protocol, the SUs access PU channels opportunistically and sense the presence of PUs periodically. The sensing period for each channel is optimized to maximize the expected throughput by minimizing~\cite[Eq. (1)]{kim_tmc_2008} which quantifies the sensing overhead (denoted by SSOH) and the missed channel access opportunities (denoted by UOPP). The optimal sensing period is derived as a function of the PU traffic parameters, specifically, the mean PU off-time, $1/\lambda_f$, and the mean PU duty cycle, $u$. We show that when the PU traffic parameters estimation error increases, the performance of the proposed MAC protocol (measured in terms of UOPP and SSOH) deteriorates. The results of the investigation on the impact of the PU traffic parameters estimation error on this MAC protocol are presented in Fig.~\ref{fig:mac_example}. We observe that as the deviations between the actual and estimated (i) mean PU off-time (Fig.~\ref{fig:mac_lambda}) and (ii) mean PU duty cycle (Fig.~\ref{fig:mac_u}) increase, the level of sensing overhead and missed opportunities, SSOH+UOPP, increase. For example, even when the estimation error in $u$ is only 15\%, the resulting SSOH+UOPP exceeds the optimal value (i.e. having perfect estimates of PU traffic parameters) by almost 10\%. Furthermore, we observe that inaccurately estimated $u$ has a more profound impact on the performance of the MAC protocol, than inaccurately estimated mean off-time.

\subsection{Related Work}
\label{sec:related_work}

A large number of algorithms in DSA systems, considering all layers of the communication stack, assume perfect knowledge of the PUs' traffic parameters, see for example~\cite[Sec. 2.1]{park_tmc_2011},~\cite[Sec. II-B]{gabran_tvt_2011},~\cite[Sec. II]{jiang_twc_2009},~\cite[Sec. III]{Chou_jsac_2007},~\cite[Sec. 3.1]{wang_tmc_2012},~\cite[Sec. 3.2]{wang_tmc_2011},~\cite[Sec. II]{wang_twc_2012},~\cite[Sec. II-A]{wang_tvt_2012},~\cite[Sec. III-A]{jung_ton_2012}. These parameters include the mean PU duty cycle, and the mean PU off-time and on-time. In reality, however, DSA systems need to periodically estimate the level of PU traffic before making any decisions on PU channel access. As DSA systems often cannot assume any a priori knowledge regarding the PU traffic parameters of the accessed channels, blind or semi-blind estimation methods of time-domain PU channel occupancy statistics need to be employed. Therefore, the issue of efficient estimation of traffic parameters of the PU, considering analytical models of the estimation process, started to gain attention from the research community. Recently published~\cite{plummer_tmc_2012} is a good example of a DSA system where the need for the most accurate estimation of PU traffic is essential. Therein, a system which scavenges spectrum opportunities in the range of (0,400] milliseconds is introduced and implemented on the TelosB mote (TI MSP40 microcontroller, Chipcon CC2420 transceiver) operating on a 2.4\,GHz range wireless sensor network testbed. One of the components of the designed channel access engine is the channel measurement and modeling component. Unfortunately, the paper does not discuss how to design such a module. Moreover, two designed channel access strategies: (i) Contiguous Secondary User Transmission and (ii) Divided Secondary User Transmission, rely strongly on the knowledge of PU traffic (denoted as ``whitespace probability density function'' by the authors). All PU traffic profiles were artificially generated beforehand and known to the channel access engine, which is an unrealistic assumption.

The most notable results dealing with analytical estimation of PU time-domain traffic parameters can be found in~\cite{kim_tmc_2008,kim_dyspan_2008,liang_ita_2010,liang_tmc_2011}. For analytical tractability, all considered works assume that PUs have exponentially distributed off- and on-times. In~\cite{kim_tmc_2008} maximum likelihood estimation was adopted for estimating the mean PU off-time while sample stream averaging was used for estimating the mean PU duty cycle. Meanwhile in~\cite{kim_dyspan_2008}, Bayesian estimation was proposed for estimating the mean PU off- and on- times. Uniform traffic sampling was assumed for both~\cite{kim_tmc_2008} and~\cite{kim_dyspan_2008}. On the other hand, the authors in~\cite{liang_ita_2010,liang_tmc_2011}, using the notion of Fisher information, derived optimal traffic sampling schemes for estimating the mean PU off-time. They argued that for a fixed channel observation window and a fixed number of samples, random sampling outperforms uniform sampling. However, perfect knowledge of the mean PU duty cycle was assumed. Besides, no closed form expressions for the accuracy of the estimated mean PU off-time was derived and different random sampling schemes were evaluated only via simulations. Unfortunately, in all aforementioned works~\cite{kim_tmc_2008,kim_dyspan_2008,liang_ita_2010,liang_tmc_2011}, the estimation accuracy, measured in terms of the mean squared error (MSE) in the estimated parameters, was not quantified in a closed form. Moreover, the impact of spectrum sensing errors on the estimation accuracy was not studied analytically. In~\cite{tehrani_tsp_2012} the authors derived the bounds on the accuracy of the joint estimation of the arrival and departure rates of PUs. However, the authors assumed that the PU traffic is observed continuously, which is an assumption that is far from being practical as the PU traffic is sampled according to a discrete sampling process.  Also, just like in earlier works, the impact of spectrum sensing errors in~\cite{tehrani_tsp_2012} was not considered.

In the context of our work we need to refer to other studies on PU traffic estimation. Specifically,~\cite{saad_arxiv_2012} followed a different approach for estimating the PU channel usage statistics (i.e. its complete distribution) by using a combination of statistical distance metrics: kernel density estimation, goodness-of-fit testing (utilizing the Kolmogorov-Smirnov test), and Kullback-Leibler distance. To increase the complexity of the problem, cooperation between spatially separated DSA nodes was considered resulting in node-to-node variances in PU traffic observations. Unfortunately, no closed-form expressions for the PU traffic distribution estimation accuracy were presented. Only a heuristic estimator (in the form of the algorithm presented in Table I of~\cite{saad_arxiv_2012}) was used. The proposed heuristic estimator is based on an example of a utility function. Moreover, the impact of spectrum sensing errors was not considered (however, errors due to fading and propagation characteristics were included).

Finally,~\cite{liu_twc_2012,wang_arxiv_2012} considered the estimation of the PU channel state through randomized channel probing. These papers modeled the PU state estimation problem as an exploration/exploitation problem and based the analysis on multi-armed bandit formulation. The difference between these two papers lies in system model assumptions and new features that have not been considered in earlier works on multi-armed bandit problems for DSA, i.e.~\cite{liu_twc_2012} considered spectrum sensing errors, while~\cite{wang_arxiv_2012} considered PU state/channel fading correlation. We need to emphasize however that PU state estimation in~\cite{liu_twc_2012,wang_arxiv_2012} has the following limiting features: (i) PU channel state estimation reduces to one parameter only (on or off time), (ii) the estimation process requires network feedback, e.g. via ACK/NACK, and (iii) the estimator does not collect statistics on the PU channel usage.

\subsection{Our Contribution}
\label{sec:contribution}

In this work, we first consider the problem of estimating the mean PU duty cycle, $u$. We derive the estimation MSE\footnote{In parameter estimation literature, the MSE is often used as a metric for the estimation accuracy for a number of reasons. The MSE is an intuitive metric that describes the average squared deviation of the estimated parameter from the actual value of the parameters. Moreover, the MSE accounts in the same manner for both positive and negative deviations. Finally, the MSE metric is mathematically tractable and can often be expressed in closed form, as opposed to the mean absolute error, or the error probability. Closed form expressions have the advantages of providing intuition regarding the results, and enabling incorporating other mathematical tools to analyze the results.} in $u$ when sample stream averaging with uniform sampling is used. We extend our work to include non-uniform sampling as well as weighted averaging with uniform sampling. Moreover, we propose estimating $u$ using maximum likelihood estimation under uniform sampling, and derive the corresponding Cram\'{e}r-Rao (CR) bound which provides a lower bound on the estimation error for unbiased estimators employing uniform sampling. Regarding the mean PU off- and on-times, we derive the CR estimation error bounds for both parameters under uniform sampling, and present the corresponding maximum likelihood estimators. All of the estimation error expressions presented in this work are formulated as functions of the total number of samples, which serves as a guideline for energy-constrained applications where the energy budget for sampling, and hence the total number of samples, is limited. We also quantify the relationship between the estimation error and the length of the observation window. This is important for delay-constrained applications, and when non-stationary traffic is considered, as it shows the compromise between the delay in learning the PU traffic parameters and the estimation error in the parameters. Besides, the effect of spectrum sensing errors on the estimation accuracy is studied analytically for the averaging estimators, while simulation results are used for the ML estimators of $u$, and the mean PU off- and on-times. Finally, we use the resulting expressions to design algorithms for the blind estimation of the PU traffic parameters under a variety of constraints, and compare their performance against the derived theoretical bounds.

The paper is organized as follows. Section~\ref{sec:System_Model} presents the system model considered in this work. Expressions for the MSE in estimating the PU traffic parameters are derived in Section~\ref{sec:squared_error} (duty cycle) and Section~\ref{sec:lambda_f_estimation} (off- and on-time rate parameters). Two practical algorithms for estimating the PU duty cycle and mean off- and on-time are presented in Section~\ref{sec:algorithm}, while numerical results are given in Section~\ref{sec:numerical_results}. Finally, Section~\ref{sec:conclusions} concludes the paper.

\section{System Model}
\label{sec:System_Model}

Following the model introduced in~\cite{kim_tmc_2008}, we consider a single channel that is licensed to a single PU\footnote{Note that this, and other assumptions of~\cite{kim_tmc_2008}, like, e.g., (i) introduction of (collaborative) spectrum sensing, (ii) listen-before-talk policy, (iii) scheduling of quiet periods, (iv) availability of the control channel, are standard and axiomatic in the DSA literature. Therefore our results are a natural extension of a well-established path in DSA research.}. The PU traffic is assumed to be stationary over a sufficiently large time window with exponentially distributed off- and on-times\footnote{We are considering a continuous model as it is more general than a discrete one, encapsulating the discrete traffic case. Furthermore, discrete PU traffic models impose an implicit synchrony between the PU and DSA networks. This requires a priori knowledge of the PU properties, e.g. guard intervals or pilot symbols. An example of such operation is in~\cite{Papadimitratos_commag_2005}, where the DSA system operates following the slot boundaries of a GSM system. To avoid such constricting requirements, we made as little assumptions on the PU properties as possible.}. The probability density function of an exponentially distributed random variable, $x$, is given as~\cite[Eq. (3.15)]{vanmieghem_book_2006} $f_{\lambda}(x)=\lambda e^{-\lambda x}$, for $x\geq 0$ and $f_{\lambda}(x)=0$, otherwise, where $\lambda$ is denoted by the rate parameter. With $\lambda=\lambda_f$ and $\lambda=\lambda_n$, $f_{\lambda}(x)$ denotes the distribution of PU off- and on-times, respectively\footnote{Note that the assumption on the exponential distribution of off- and on-times is common in DSA literature, e.g., see recent examples of~\cite{li_jsac_2011,alshamrani_jsac_2011,zhang_jsac_2011}; see also recent papers confirming the exponential distribution of time-domain utilization of certain licensed channels~\cite{wellens_phycom_2009,chen_mobicom_2009,yin_tmcsub_2010}.}. The mean PU off- and on-times are equal to the reciprocal of $\lambda_f$ and $\lambda_n$, respectively. Besides, the duty cycle $u$ of the PU can be calculated as~\cite[Sec. 11.3]{vanmieghem_book_2006} $u =\frac{\lambda_f}{\lambda_f+\lambda_n}$. Hence, $\lambda_f$, $\lambda_n$ and $u$ are inter-dependent, where estimating any two of the three parameters is sufficient to completely estimate the PU traffic parameters.

In order to estimate the traffic parameters, the PU channel is sampled in order to acquire data regarding the state of the PU (on or off). For the considered system model, denote the total number of samples by $N$. Denote the PU traffic samples by the vector $\boldsymbol{z} = \left[z_1, z_2, \cdots, z_N\right]$ where $z_n$ is the $n$th traffic sample, and $z_n=1$ if the PU is active and $z_n=0$, otherwise. Moreover, in the proposed model, we consider the general case where the spectrum sensing process is prone to errors. The sensing error is modeled in the form of false alarm and mis-detection probabilities, denoted by $P_f$ and $P_m$, respectively. The sensing error is assumed to be independent for different traffic samples. The estimated PU traffic samples are denoted by the vector $\boldsymbol{\tilde z} = \left[\tilde z_1, \tilde z_2, \cdots, \tilde z_N\right]$ where $\tilde z_n$ is the $n$th estimated traffic sample. It follows that $\tilde z_n=1$ if $z_n=1$ and no mis-detection error occurred, or $z_n=0$ and a false alarm error occurred. Similarly, $\tilde z_n=0$ if $z_n=1$ and a mis-detection error occurred, or $z_n=0$ and no false alarm error occurred. Furthermore, the inter-sample times are given by the vector $\mathcal{T}=\left[T_1,T_2,\cdots,T_{N-1}\right]$ where $T_n$ denotes the time between samples $z_n$ and $z_{n+1}$. Finally, the total observation window length is denoted by $T$, where $T = \sum_{n=1}^{N-1} T_n$.

Denote the PU state transition probability by $\Pr_{xy}(t)$, which corresponds to the probability that the PU state changes from state $x$ to state $y$ within time $t$, where $\{x,y\}=0$ denotes that the PU is idle while $\{x,y\}=1$ denotes that the PU is active. The PU state transition probabilities were derived in~\cite[Sec. 6.1]{kim_tmc_2008} as
\begin{subnumcases}{\textstyle\Pr_{xy}(t)=\label{eq:p_xy}}
1-u+u e^{\frac{-\lambda_f t}{u}}, & $x=0,y=0$,\\
1-\textstyle\Pr_{00}(t), & $x=0,y=1$,\\
u+(1-u) e^{\frac{-\lambda_f t}{u}}, & $x=1,y=1$,\\
1-\textstyle\Pr_{11}(t) & $x=1,y=0$.
\end{subnumcases}
In this work $\Pr_{xy}(t)$ is later used to derive the MSE in the estimates of $u$, $\lambda_f$, and $\lambda_n$.

As remarked in Section~\ref{sec:Introduction}, estimators of $u$ and $\lambda_f$ are analytically described in closed form in~\cite{kim_tmc_2008,kim_dyspan_2008,liang_ita_2010,liang_tmc_2011}. However, a measure of the estimation error in $u$ and $\lambda_f$, was not given, noting that in~\cite[Sec. 6.2]{kim_tmc_2008} only the asymptotic confidence interval for the estimates of $u$ and $\lambda_f$ was presented. In the following sections, we propose new methods to estimate $u$ and we derive the MSE in the estimates of $u$, $\lambda_f$, and $\lambda_n$.

\section{Estimation of the Primary User Duty Cycle $u$}
\label{sec:squared_error}

In this section, we analyze different methods for estimating the duty cycle, $u$, of the PU. We first present an estimator based on averaging the traffic samples, labeled the \emph{averaging estimator}, similar to the estimator presented in~\cite{kim_tmc_2008,kim_dyspan_2008,liang_ita_2010,liang_tmc_2011}. In addition, we modify the estimator to the general case where the PU traffic samples are not uniformly sampled. Furthermore, as we observe that the estimation accuracy is bounded by the sample correlation, we propose two different estimation methods to alleviate the correlation effect. The first method is based on the weighted averaging of the traffic samples, labeled the \emph{weighted averaging estimator}, and the second method is based on maximum likelihood (ML) estimation. For all three estimation methods, we derive expressions for the MSE in the estimates. Moreover, we derive the CR bound on the estimation error when using uniformly sampled traffic samples. The MSE expressions are presented as functions of the number of samples and the observation window length to serve as guidelines for traffic estimation in energy-constrained and delay-constrained systems, respectively.

\subsection{The Averaging Estimator under Perfect Sensing}
\label{sec:estimation_u}

The averaging duty cycle estimator, $\tilde{u}_a$, is defined as ~\cite[Sec. 6.1]{kim_tmc_2008}
\begin{equation}
\tilde{u}_a = \frac{1}{N} \sum_{n=1}^{N} \tilde z_n.
\label{eq;u_est}
\end{equation}
We first consider the case where the spectrum sensing errors can be ignored, i.e., $P_f = P_m = 0$, hence, $\tilde z_n = z_n \forall n$. The impact of spectrum sensing errors on the estimation error is presented in the next section.

\subsubsection{The MSE in $\tilde{u}_a$}
\label{sec:mse_u_non_uni}

The MSE in $\tilde{u}_a$ for $N$ samples can be defined as
\begin{equation}
V_{\tilde{u}_a,N} = E\left[\tilde{u}_a^2\right]-u^2,
\label{eq;u_var_0}
\end{equation}
where $E[\cdot]$ denotes the expectation. The intuition behind (\ref{eq;u_var_0}) is as follows; the expectation is calculated over all possible values of $\tilde{u}_a$ resulting from all $2^N$ permutations of the traffic samples vector $\boldsymbol{z}$. Define $\mathcal{Z}$ as a vector containing all $2^N$ permutations of $\boldsymbol{z}$ with $\mathcal{Z}_n$, $n \in \{1,2,\cdots,2^N\}$, defined as the $n$th element of $\mathcal{Z}$. Furthermore, define $\mathcal{Z}_{n,m}$, $m \in \{1,2,\cdots,N\}$, as the $m$th traffic sample of $\mathcal{Z}_n$, and define $\zeta_n = \sum_{m=1}^{N} \mathcal{Z}_{n,m}$, i.e., the summation of all traffic samples of $\mathcal{Z}_n$. Then, substituting (\ref{eq;u_est}) in (\ref{eq;u_var_0}) yields
\begin{equation}
V_{\tilde{u}_a,N} = \frac{1}{N^2}\sum_{n=1}^{2^N} \zeta_n^2 \Pr(\boldsymbol{z} = \mathcal{Z}_n|\mathcal{T})-u^2,
\label{eq;u_var_1}
\end{equation}
where $\Pr(\boldsymbol{z} = \mathcal{Z}_n|\mathcal{T})$ denotes the probability of observing PU traffic sample sequence $\mathcal{Z}_n$, for a given vector of sampling times $\mathcal{T}$. We then have the following corollary.
\begin{corollary}
The MSE of $\tilde{u}_a$ is given as
\begin{equation}
V_{\tilde{u}_a,N} = \frac{u(1-u)}{N} + \frac{2u(1-u)}{N^2} \sum_{i=1}^{N-1} \sum_{j=1}^{N-i} \prod_{k=j}^{i+j-1} e^{\frac{-T_{k}\lambda_f}{u}},
\label{eq;nb_1}
\end{equation}
\begin{IEEEproof}
See the proof of (\ref{eq;nb_1_se}).
\end{IEEEproof} 
\end{corollary}
From Corollary 1 we obtain the subsequent corollary.
\begin{corollary}
\label{cor_3}
The decrease in the MSE in $\tilde{u}_a$ with each extra sample, $D_{\tilde{u}_a,N+1}$, is given as
\begin{align}
D_{\tilde{u}_a,N+1} & = V_{\tilde{u}_a,N} - V_{\tilde{u}_a,N+1} \nonumber \\ &= \frac{(2N+1)}{(N+1)^2}V_{\tilde{u}_a,N} - \frac{u(1-u)}{(N+1)^2}\nonumber\\&\quad\times \left(1+2\sum_{i=0}^{N-1} \prod_{k=N-i}^{N} e^{\frac{-T_{k}\lambda_f}{u}} \right).
\label{eq;vu_nu_dec}
\end{align}
\begin{IEEEproof}
Via elementary algebra.
\end{IEEEproof} 
\end{corollary}
Corollary~\ref{cor_3}, as we will show in Section~\ref{sec:algorithm}, proves important in designing adaptive algorithms for the blind estimation of $u$.

\paragraph*{Remarks}

The rightmost term of (\ref{eq;nb_1}) represents the increase in the estimation error caused by the correlation between the traffic samples. As $T_k$ tends to infinity, this term tends to zero, hence, $V_{\tilde{u}_a,N}$ approaches $\frac{u(1-u)}{N}$, which is the MSE in estimating the duty cycle of an uncorrelated traffic sample sequence\footnote{Note that $\frac{u(1-u)}{N}$ is the variance of a binomial distribution normalized by $N^2$ where the probability of success is set to $u$~\cite[Ch. 4]{prob_schaum}.}. This is attributed to the fact that the inter-sample time becomes large compared to the mean off- and on-times of the PU, hence, the correlation between the samples vanishes. The estimation error is a function of the traffic parameters, the number of samples, and the inter-sample time sequence. The optimal inter-sample time sequence that minimizes the estimation error for a given number of samples and a fixed total observation window length is derived in the next section.

\subsubsection{The Optimal Inter-Sample Time Sequence for Minimizing the MSE in $\tilde{u}_a$}
\label{sec:optim_u}

In this section, the MSE in $\tilde{u}_a$ is shown to be convex with respect to the inter-sample time sequence, $\mathcal{T}$. The optimal $\mathcal{T}$, denoted by $\mathcal{T}^*$, is derived, and the corresponding expression for the MSE in $\tilde{u}_a$ is presented. Expression (\ref{eq;nb_1}) is proven to be convex by showing that the Hessian of $V_{\tilde{u}_a,N} \left( \mathcal{T} \right)$, denoted by $\nabla^2 V_{\tilde{u}_a,N} \left( \mathcal{T} \right)$, is positive-semidefinite~\cite{berghe}. The proof of convexity is given in Appendix~\ref{sec:V_u_conv}. 

The problem of minimizing $V_{\tilde{u}_a,N} \left( \mathcal{T} \right)$ with respect to $\mathcal{T}$ can be written as:
\begin{align}
\text{minimize } & V_{\tilde{u}_a,N} (\mathcal{T}) = \frac{u(1-u)}{N} + \frac{2u(1-u)}{N^2}\nonumber\\&\qquad\qquad\qquad\times \sum_{i=1}^{N-1} \sum_{j=1}^{N-i} \prod_{k=j}^{i+j-1} e^{-\frac{T_{k}\lambda_f}{u}};\\
\text{subject to } & -T_{n} \leq 0, n = 1,2,\cdots,N-1 \label{eq:opt1};\\ 
& \sum_{n=1}^{N-1} T_{n} = T. \label{eq:opt2}
\end{align}
The optimization problem can be solved by Lagrangian duality~\cite{berghe} where the Lagrangian function can be expressed as
\begin{align}
L_V\left(\mathcal{T},\boldsymbol{\upsilon},\mu\right)& = V_{\tilde{u}_a,N}\left( \mathcal{T} \right) - \sum_{k = 1}^{N-1} \upsilon_k T_k\nonumber\\&\qquad + \mu \left(\sum_{k=1}^{N-1} T_{k} - T\right),
\end{align}
where $\boldsymbol{\upsilon}=[\upsilon_1,\upsilon_2,\cdots,\upsilon_{N-1}]$ is the vector of the Lagrangian multipliers associated with inequalities (\ref{eq:opt1}) and $\mu$ is the Lagrangian multiplier associated with (\ref{eq:opt2}). As the optimization problem is convex, and the objective and constraint functions are differentiable, the optimal inter-sample time sequence, $\mathcal{T}^*$, satisfies the Karush-Kuhn-Tucker (KKT) conditions:
\begin{subnumcases}{}
-T_n^* \leq 0, \label{eq:KKT3} \\
\sum_{k=1}^{N-1} T_k^* - T = 0, \label{eq:KKT2} \\
\upsilon_n^* T_n^* = 0 , \upsilon_n^* \geq 0 , \label{eq:KKT4} \\
\nabla V_{\tilde{u}_a,N}\left(T_n^*\right) - \upsilon_n^* + \mu^* = 0, \label{eq:KKT1}
\end{subnumcases}
where $n \in \{1,2\cdots,N-1\}$ and the superscript $*$ signifies optimality. Expression (\ref{eq:KKT1}) can be expressed as $\frac{-2u\left(1-u\right)\lambda_f}{N^2u} \sum_{i=1}^{n} \sum_{j=n}^{N-1} \prod_{k=i}^{j} e^{-\lambda_f T_k^*/u} - \upsilon_n^* + \mu^* = 0$. The solution of the convex problem is first presented for the special case where $T_n^*>0$, $\upsilon_n^*=0 \text{ } \forall n$, i.e.\ when all samples have non-zero spacing in time. The solution is later expanded to include cases where $T_n^*=0$ for a set of $n$, i.e.\ samples coincide in time implying that the sample is weighted by the number of coinciding samples. 

For the case where $T_n^*>0 \text{ } \forall n$, $\mathcal{T}^*$ is given as follows. Denote $\Gamma_a^* = e^{\frac{-\lambda_f T_a^*}{u}}$ and $\Gamma_b^* = e^{\frac{-\lambda_f T_b^*}{u}}$, then $T_n^*=T_a^*$ for $n = 1$ and $n=N-1$, and $T_n^* = T_b^*$, otherwise, where
\begin{subnumcases}{}
\Gamma_a^* = \frac{\Gamma_b^*}{1-\Gamma_b^*}, \label{eq;opt_sol1}\\
2T_a^* + \left(N-3\right)T_b^* = T, \label{eq;opt_sol2}\\
T > \left(N-3\right)\frac{u}{\lambda_f}\log 2. \label{eq;opt_sol3}
\end{subnumcases}
Equation (\ref{eq;opt_sol1}) is derived by simultaneously solving (\ref{eq:KKT1}) for $n= 1$ and $n=2$. Equation (\ref{eq;opt_sol2}) is derived from condition (\ref{eq:KKT2}). Condition (\ref{eq;opt_sol3}) is derived by setting $T_a^*=0$ in (\ref{eq;opt_sol1}) and (\ref{eq;opt_sol2}) and solving for $T$. Expressions (\ref{eq;opt_sol1}) and (\ref{eq;opt_sol2}) can be shown to satisfy (\ref{eq:KKT3})--(\ref{eq:KKT1}) but the proof is omitted for brevity. The solution for the optimization problem implies that as the length of the total observation window, $T$, increases, the optimal inter-sample time sequence approaches uniform sampling. As $T$ decreases, $\mathcal{T}^*$ remains uniform for samples $2$ to $N-1$. However, the first and last inter-sample times are equal in length, and shorter in length than the rest of the inter-sample times. If $T$ is decreased to $\frac{\left(N-3\right)u}{\lambda_f}\log 2$, $T_1^*$ and $T_{N-1}^*$ approach zero, i.e.\ the first two samples and last two samples coincide. This implies that the number of samples is decreased to $N-2$ and the first and last samples are weighted by two.

For the case where $T \leq \frac{\left(N-3\right)u}{\lambda_f}\log 2$, $T_n^*$ can be equal to zero for $n \in \mathcal{K}$ where $\mathcal{K} = \{1,2,\cdots,k-2,k-1, N-k+1, N-k+2,\cdots,N-1\}$ and $1<k<\lfloor \frac{N}{2} \rfloor$. $\mathcal{T}^*$ can be found by solving (\ref{eq:KKT3})--(\ref{eq:KKT1}) where $T_n^* = 0$ for $n \in \mathcal{K}$ and $T_n^* > 0, \upsilon_n^* = 0$, otherwise. $\mathcal{T}^*$ is derived as $T_n^*=0$ for $n \in \mathcal{K}$, $T_n^*=T_a^*$ for $n = k$ and $n=N-k$ and $T_n^* = T_b^*$, otherwise, where
\begin{subnumcases}{}
\Gamma_a^* = \frac{\Gamma_b^*}{k\left(1-\Gamma_b^*\right)},\label{eq;opt_sol_gen_1}\\
2T_a^* + \hat{N}T_b^* = T,\label{eq;opt_sol_gen_2}\\
\hat{N}\frac{u}{\lambda_f} \log \frac{k+1}{k} < T \leq \hat{N} \frac{u}{\lambda_f}\log \frac{k}{k-1},\label{eq;opt_sol_gen_3}
\end{subnumcases}
where $\hat{N}\triangleq N-2k-1$. Equation (\ref{eq;opt_sol_gen_1}) is derived by simultaneously solving (\ref{eq:KKT1}) for $n= k$ and $n=k+1$. Equation (\ref{eq;opt_sol_gen_2}) is derived from condition (\ref{eq:KKT2}). The lower bound in (\ref{eq;opt_sol_gen_3}) is derived by setting $T_a^*=0$ in (\ref{eq;opt_sol_gen_1}) and (\ref{eq;opt_sol_gen_2}) and solving for $T$. The upper bound in (\ref{eq;opt_sol_gen_3}) is based on the condition $\upsilon_{k-1}^* \geq 0$ where the expression for $\upsilon_n^*$ can be derived for $n \in \mathcal{K}$ from (\ref{eq:KKT1}), using (\ref{eq;opt_sol_gen_1}), as $\upsilon_n^* = \frac{2u\left(1-u\right)\lambda_f}{N^2u}\left(k-n\right)\left[\frac{\Gamma_a^*}{1-\Gamma_b^*}-n\right]$. Again, (\ref{eq;opt_sol_gen_1}) and (\ref{eq;opt_sol_gen_2}) satisfy (\ref{eq:KKT3})--(\ref{eq:KKT1}) $\forall n$, and the proof is omitted for brevity. The solution for the optimization problem implies that if $T$ falls in the boundary expressed in (\ref{eq;opt_sol_gen_3}), then the first and last $k-1$ samples are omitted, and the $k$th and $\left(N-k\right)$th samples are weighted by $k$. Again, the middle $N-2k-1$ inter-sample times are uniformly sampled and $T_k^* = T_{N-k}^*$.

For the special case where $0 < T \leq (N-2k+1) \frac{u}{\lambda_f} \log \frac{k}{k-1}$ and $k = \lfloor \frac{N}{2} \rfloor$, all inter-sample times decrease to zero except for the middle interval, for $N$ even, or the middle two intervals, for $N$ odd. Accordingly, for even $N$, $T_n^* = 0$ for $n\neq \frac{N}{2}$ and $T_{N/2}^* = T$, whereas for odd $N$, $T_n^* = 0$ for $n \notin \{\frac{N-1}{2},\frac{N+1}{2}\}$ and $T_{\frac{N-1}{2}}^*=T_{\frac{N+1}{2}}^*=T/2$. This can be shown to satisfy the KKT conditions (\ref{eq:KKT3})--(\ref{eq:KKT1}).

Using the optimal inter-sample time sequence, the lower bound on the estimation error for the averaging estimator, $\tilde{u}_a$, can be derived by substituting (\ref{eq;opt_sol_gen_1}) and (\ref{eq;opt_sol_gen_2}) in (\ref{eq;nb_1}) yielding
\begin{align}
V_{\tilde{u}_a,N}^* &= \frac{2u(1-u)}{N^2}\nonumber\\&\qquad\times\left( \frac{N}{2} + \frac{\Gamma_b^* + k(k-1)(1-\Gamma_b^*)^2}{(1-\Gamma_b^*)^2}\right.\nonumber\\&\left.\qquad + \frac{\Gamma_b^*(N-2k)(1-\Gamma_b^*)}{(1-\Gamma_b^*)^2}\right),
\label{eq;low_bound_non_uni}
\end{align}
for $0<k<\lfloor \frac{N}{2} \rfloor$ where $k$ is chosen to satisfy (\ref{eq;opt_sol_gen_3}) and $\Gamma_b^*$ is found by solving (\ref{eq;opt_sol_gen_1}) and (\ref{eq;opt_sol_gen_2})\footnote{For $k = \lfloor \frac{N}{2} \rfloor$, for even $N$, $V_{\tilde{u}_a,N}^* = \frac{2u(1-u)}{N^2}\left( \frac{N}{2} + \left(N^2(e^{-\lambda_f T/u}+1)-2N\right)/4\right)$, and for odd $N$, $V_{\tilde{u}_a,N}^* = \frac{2u(1-u)}{N^2}\left( \frac{N}{2} + (N-1)\left((N-1)e^{-\lambda_f T/u}\right.\right.$ $\left.\left.+4e^{-\lambda_f T/(2u)}+(N-3)\right)/4\right)$.}.

\paragraph*{Remarks}
\label{sec:u_av_estimation_Remarks}

The optimal inter-sample time sequence, $\mathcal{T}^*$, is non-uniform where the first and last inter-sample times are shorter than the rest of the inter-sample times. Moreover, generally, the first and last samples have higher weights compared to the rest of the samples. This is attributed to the fact that the first and last samples are at the edges of the traffic sample sequence, hence, they have lower correlation with the rest of the traffic samples. However, as the total observation window length increases, the impact of the first and last samples on the overall estimation error decreases, and the gap between the estimation error for the optimal non-uniform sampling sequence and the uniform sampling sequence diminishes. Furthermore, $\mathcal{T}^*$ is a function of $u$, the very same parameter that is to be estimated, as well as a function of the mean PU departure rate, $\lambda_f$ (or the mean PU arrival rate, $\lambda_n$, as $\lambda_f$ equals $u\lambda_n/(1-u)$), which is not necessarily known by the traffic estimator. Hence, $\mathcal{T}^*$ cannot be known a priori, yet $\mathcal{T}^*$ can be used as a guideline in algorithm design for the blind estimation of the traffic parameters. For instance, apart from `weighting' the first and last samples, $\mathcal{T}^*$ is found to be an almost uniformly sampled sequence. Besides, the error expression given in (\ref{eq;low_bound_non_uni}) serves as a lower bound on the MSE in estimating $u$ using averaging for any inter-sample time sequence.

\subsubsection{The Averaging Estimator under Uniform Sampling}
\label{sec:mse_u}

The work in~\cite{kim_tmc_2008,kim_dyspan_2008,liang_ita_2010,liang_tmc_2011} considered estimating $u$ by averaging uniformly sampled traffic observations. In this section, we derive the MSE in $\tilde{u}_a$ under uniform sampling, denoted by $V_{\tilde{u}_{ua},N}$. With constant inter-sample times, $T_n = \frac{T}{N-1} = T_u$, $\forall T_n \in \mathcal{T}$. Substituting in (\ref{eq;nb_1}), the MSE can be written as
\begin{align}
V_{\tilde{u}_{ua},N} &= \frac{2u(1-u)}{N^2}\left( \frac{N}{2} + \sum_{i=1}^{N-1} \Gamma_u^i(N-i)\right)\nonumber\\ & = \frac{2u(1-u)\Gamma_u(\Gamma_u^N-N(\Gamma_u-1)-1)}{N^2(1-\Gamma_u)^2}\nonumber\\&\qquad + \frac{u(1-u)}{N},
\label{eq;uni_samp_u}
\end{align}
where $\Gamma_u = e^{-\frac{\lambda_f T_u}{u}}$. Again, the leftmost part in the second equation of (\ref{eq;uni_samp_u}) accounts for the increase in estimation error caused by sample correlation. Intuitively, when the sample correlation is high, increasing $N$ in a fixed time window leads to an insignificant change in the estimation error. Formally, we obtain the following corollary.
\begin{corollary}
\label{th:v_ul}
For a fixed observation window length, as the number of samples increases, the MSE error in estimating $u$ for uniform sampling approaches an asymptote $V_{\tilde{u}_{ua},L}$, where
\begin{equation}
V_{\tilde{u}_{ua},L} = \lim_{N\to\infty} V_{\tilde{u}_{ua},N} = \frac{2u(1-u)}{\eta^2}\left(e^{-\eta}+\eta-1\right),
\label{eq:nb_2}
\end{equation}
where $\eta = \frac{T\lambda_f}{u}$.
\begin{IEEEproof}
Via elementary algebra.
\end{IEEEproof}
\end{corollary}

Note that $V_{\tilde{u}_{ua},L}$ tends to 0 as the observation window length is increased. Using Corollary \ref{th:v_ul}, the number of samples, $N$, can be chosen such that the resulting error is above the asymptotic error (\ref{eq:nb_2}) by a factor $\beta$. Then $N$ can be evaluated by solving $V_{\tilde{u}_{ua},N} = \beta V_{\tilde{u}_{ua},L}$.

\paragraph*{Remarks}

When estimating $u$ by averaging uniformly sampled traffic samples, the estimation error is lower bounded. The lower bound is caused by sample correlation and can only be eliminated by increasing the total observation window length.

\subsection{The Averaging Estimator under Imperfect Sensing}
\label{sec:estimation_u_err}

The analysis presented in Section~\ref{sec:estimation_u} is extended here to include the effect of spectrum sensing errors on the estimation error. Introducing spectrum sensing errors to the averaging estimator expressed in (\ref{eq;u_est}) causes the estimator to become biased. The expected value of the estimator can be calculated as $E\left[\tilde{u}_a\right] = P_f\left(1 - u\right) + u\left(1 - P_m\right)$, where the expectation is calculated over all possible values of $\tilde{u}_a$ resulting from all $2^N$ permutations of the estimated PU traffic samples vector $\boldsymbol{\tilde z}$. Thus, the duty cycle can be calculated from $E\left[\tilde{u}_a\right]$ where $u = (E\left[\tilde{u}_a\right]-P_f)/(1-P_f-P_m)$. Accordingly, we define the following unbiased estimator\footnote{Note that the estimator $\tilde{u}_{a,s}$ is not defined for the special case of $P_f + P_m = 1$. For $P_f + P_m = 1$, the denominator of the proposed unbiased estimator equals zero, and the expectation of the biased estimator can be expressed as $E\left[\tilde{u}_a\right] = P_f$, which is independent of $u$. Hence, both estimators fail to estimate $u$. On the other hand, note that $P_f+P_m\geq1$ does not correspond to any relevant practical sensing method. Typical values for the probability of false alarm and mis-detection are $P_f\lessapprox0.1$ and $P_m\lessapprox0.1$, respectively, e.g.,~\cite[Sec. VII-C]{arslan_tutorial},~\cite[Sec. VI-A]{tian_twc_2012},~\cite{Gerihofer_commag_2007,stevenson_commag_2009},~\cite[Sec. 6.6]{min_tmc_2011},~\cite[Sec. IV-A]{kim_twc_2010}.}
\begin{equation}
\tilde{u}_{a,s} = \frac{1}{1-P_f-P_m}\left[-P_f + \frac{1}{N} \sum_{n=1}^{N} \tilde{z}_n\right],
\label{eq;u_est_1}
\end{equation}
$\forall P_f, P_m: P_f+P_m\neq1$. Define $\mathcal{\tilde Z}$ as a vector containing all $2^N$ permutations of $\boldsymbol{\tilde z}$ with $\mathcal{\tilde Z}_n$, $n \in \{1,2,\cdots,2^N\}$, defined as the $n$th element of $\mathcal{\tilde Z}$. Furthermore, define $\mathcal{\tilde Z}_{n,m}$, $m \in \{1,2,\cdots,N\}$, as the $m$th traffic sample of $\mathcal{\tilde Z}_n$. Thus, the MSE in $\tilde{u}_{a,s}$ for $N$ samples, denoted by $V_{\tilde{u}_{a,s},N}$ can be expressed as
\begin{equation}
V_{\tilde{u}_{a,s},N} = \sum_{n=1}^{2^N} S_n^2 \Pr(\tilde{\boldsymbol{z}} = \tilde{\mathcal{Z}}_n|\mathcal{T})-u^2,
\label{eq;u_var_1a}
\end{equation}
where $S_n = \frac{1}{1-P_f-P_m}\left[-P_f + \frac{1}{N} \sum_{m=1}^{N} \tilde{\mathcal{Z}}_{n,m}\right]$. We then have the following theorem.
\begin{theorem}
The MSE in $\tilde{u}_{a,s}$ is given as
\begin{align}
V_{\tilde{u}_{a,s},N} & = \frac{2u(1-u)}{N^2} \sum_{i=1}^{N-1} \sum_{j=1}^{N-i} \prod_{k=j}^{i+j-1} e^{\frac{-T_{k}\lambda_f}{u}}+ \frac{u(1-u)}{N}\nonumber\\&\qquad + \frac{uP_m\left(1-P_m\right) + \left(1-u\right)P_f\left(1-P_f\right)}{N\left(1-P_f-P_m\right)^2},
\label{eq;nb_1_se}
\end{align}
$\forall P_f, P_m: P_f+P_m\neq1$.
\begin{IEEEproof}
See Appendix~\ref{sec:math_ind_vu}.
\end{IEEEproof} 
\end{theorem}

\paragraph*{Remarks}

Comparing (\ref{eq;nb_1_se}) with (\ref{eq;nb_1}), it is clear that the rightmost term of the right hand side of (\ref{eq;nb_1_se}) models the increase in the estimation error caused by the spectrum sensing errors. Moreover, the leftmost term of the right hand side of (\ref{eq;nb_1_se}) accounts for the estimation error caused by the sample correlation. Furthermore, unlike the impact of sample correlation on the estimation error, the effect of the spectrum sensing errors on the estimation error can be asymptotically eliminated by increasing $N$. Besides, the increase in the estimation error attributed to the spectrum sensing errors is not a function of the inter-sample time sequence, $\mathcal{T}$. Accordingly, $V_{\tilde{u}_{a,s},N}$ is convex with respect to $\mathcal{T}$, and the optimal $\mathcal{T}$ that minimizes the MSE in $\tilde{u}_{a,s}$ is the same as $\mathcal{T}$ that minimizes the MSE in $\tilde{u}_{a}$ that is derived in Section~\ref{sec:optim_u}.

\subsection{The Weighted Averaging Estimator under Perfect Sensing}

In the previous sections, the PU duty cycle, $u$, is estimated using equal weight averaging of the channel samples. The optimal inter-sample times were found to reach zero for some samples implying that weighting might improve the estimation accuracy. In this section, we propose a new estimator that averages weighted traffic samples to decrease the estimation error by alleviating the effect of sample correlation. For analytical tractability, uniform sampling is assumed with a constant inter-sample time denoted by $T_c$.

We first present the special case where the spectrum sensing errors can be neglected, that is, $P_f = P_m = 0$, and $\tilde z_n = z_n \forall n$. The effect of spectrum sensing errors on the estimation error is investigated in the next section. The estimator is defined as $\tilde{u}_w=\sum\limits_{i=1}^{N}w_{i}z_i$, where $w_i$ is the weight of sample $z_i$. Then $E[\tilde{u}_w]=u\sum\limits_{i=1}^{N}w_{i}$, thus, for the estimator to be unbiased, that is $E[\tilde{u}_w]=u$, the weights must satisfy the condition $\sum\limits_{i=1}^{N}w_{i}=1$.

\subsubsection{The MSE in $\tilde{u}_w$}
\label{sec:mse_u_weight}

The MSE in $\tilde{u}_w$ can be written as
\begin{align}
V_{\tilde{u}_w,N} & = E\left[(\tilde{u}_w-E[\tilde{u}_w])^2\right]\nonumber\\ & = E\left[\left(\sum\limits_{i=1}^{N}w_{i}z_{i}-E\left[\sum_{i=1}^{N}w_{i}z_{i}\right]\right)^{2}\right] \notag \\ & = E\left[\left(\sum\limits_{i=1}^{N}w_{i}z_{i}\right)^{2}\right]-u^{2}\left(\sum\limits_{i=1}^{N}w_{i}\right)^{2}\nonumber\\ & = u\sum\limits_{i=1}^{N}w_{i}^{2}+2\sum_{i<j}w_{i}w_{j}E[z_{i}z_{j}]\nonumber\\&\qquad-u^{2} \left(\sum\limits_{i=1}^{N}w_{i}\right)^{2}.
\label{eqn:vwu}
\end{align}
The expression $E[z_{i}z_{j}]$ in (\ref{eqn:vwu}) represents the correlation between $z_{i}$ and $z_{j}$ denoted by $R_{i,j}$. Consider $R_{i,i+j}$, $\forall j\geq1$, then
\begin{align}
R_{i,i+j} & = E[z_{i}z_{i+j}] = \Pr\{z_{i}=1, z_{i+j}=1\}\nonumber\\ & = \Pr\{z_{i}=1, z_{i+j-1}=1\}\textstyle\Pr_{11}(T_{c}) \nonumber \\ & \qquad + \Pr\{z_{i}=1, z_{i+j-1}=0\}\textstyle\Pr_{01}(T_{c})\nonumber\\ & = R_{i,i+j-1} \textstyle\Pr_{11}(T_{c})\nonumber\\&\qquad + (u-R_{i,i+j-1}) \textstyle\Pr_{01}(T_{c}).
\label{eqn:recursive}
\end{align}
The initial condition for the recursive equation (\ref{eqn:recursive}) is $R_{i,i}=E[z_{i}z_{i}]=u$. Thus, solving equation (\ref{eqn:recursive}) yields $R_{i,i+j} = u\Gamma_c^{j}+u^{2}(1-\Gamma_c^{j})$, where $\Gamma_c=e^{\frac{-\lambda_f T_c}{u}}$. Since the traffic samples have a constant mean and the correlation function is only related to the time difference between the samples, the samples follow a wide-sense stationary process\footnote{It is stated in~\cite{kim_tmc_2008} that the traffic samples follow a semi-Markov process. But given the condition of using a constant inter-sample time, it turns out to be a wide-sense stationary process.} and $R_{i,i+j} = R[j], \forall j\geq0$. Substituting $R_{i,j}$ in (\ref{eqn:vwu}) yields
\begin{align}
V_{\tilde{u}_w,N} & = u\sum\limits_{i=1}^{N}w_{i}^{2}+2\sum\limits_{j=1}^{N-1}\sum\limits_{i=1}^{N-j}w_{i}w_{i+j}\left[u\Gamma_c^{j}+u^{2}(1-\Gamma_c^{j})\right]\nonumber\\&\qquad - u^{2}\left(\sum\limits_{i=1}^{N}w_{i}\right)^{2} \nonumber \\ & = \left(\sum\limits_{i=1}^{N}w_{i}^{2}+2\sum\limits_{j=1}^{N-1}\Gamma_c^{j}\sum\limits_{i=1}^{N-j}w_{i}w_{i+j}\right)\nonumber\\&\qquad\times u(1-u).
\label{eqn:v_wu_final}
\end{align}
Note that (\ref{eqn:v_wu_final}) matches (\ref{eq;uni_samp_u}) if constant weighting is assumed, i.e., $w_{i}=1/N, i \in\{1,2,\cdots,N\}$.

\subsubsection{The Optimal Weighting Sequence}
\label{sec:opt_weight_seq}

The optimal weighting sequence that minimizes the MSE in $\tilde{u}_w$, denoted by $\mathbf{w}^* = [w_{1}^*, w_{2}^*,\cdots,w_{N}^*]^{T}$, is derived in this section. According to the orthogonality principle, $\mathbf{w}^*$ satisfies $E[(u-\tilde{u}_w)z_{j}]=0, \forall j \in\{1,2,\cdots,N\}$. Hence, $\mathbf{w}^*=\mathbf{R}^{-1}\mathbf{q}_{uz}$, where $\mathbf{R}=[\mathbf{r}_{i,j}]_{N\times N}$, $\mathbf{r}_{i,j}=R[|i-j|]$, and $\mathbf{q}_{uz}$ is the cross correlation vector $E[u\boldsymbol{z}]$. Since the traffic samples follow a wide-sense stationary process, $\mathbf{R}$ is both symmetric and Toeplitz. The cross correlation vector, $\mathbf{q}_{uz}$, can be written as $\mathbf{q}_{uz}=E[u\boldsymbol{z}]=u(E[z_1], E[z_2],\cdots,E[z_N])^T=[u^2, u^2,\cdots,u^2]^{T}$. Normalizing the weights to get an unbiased estimator, the optimal weighting sequence is given by $\mathbf{w}^*=\frac{\mathbf{R}^{-1}\mathbf{q}_{uz}}{\mathbf{c}^{T}\mathbf{R}^{-1}\mathbf{q}_{uz}}$, where $\mathbf{c}$ is a constant vector $[1, 1,\cdots,1]^{T}\in\mathbb{R}^{N\times 1}$. Hence, the optimal weighting sequence can be expressed as $w_{1}^* = w_{N}^* = \frac{1}{N(1-\Gamma_c)+2\Gamma_c}$ and $w_{i}^* = \frac{1-\Gamma_c}{N(1-\Gamma_c)+2\Gamma_c} \text{ } \forall i \notin \{1,N\}$.

The MSE in $\tilde{u}_w$ when the optimal weighting sequence is used can be found by substituting $\mathbf{w}^*$ in (\ref{eqn:v_wu_final}) yielding $V_{\tilde{u}_w,N}^* = \frac{u(1-u)(1+\Gamma_c)}{N(1-\Gamma_c)+2\Gamma_c}$. Due to the correlation between the samples, for a given observation window length, the MSE in $\tilde{u}_w$ approaches an asymptote, denoted by $V_{\tilde{u}_w,L}^*$, as $N$ increases. $V_{\tilde{u}_w,L}^*$ can be calculated as follows
\begin{equation}
V_{\tilde{u}_w,L}^* = \lim_{N\to\infty} V_{\tilde{u}_w,N}^* = \frac{u(1-u)}{1+\frac{\lambda_f T}{2u}}.
\label{eq;MSE_w_inf}
\end{equation}
\paragraph*{Remarks}

The derived optimal weighting sequence dictates that the first and last samples have to be multiplied by higher weights than the rest of the samples. This is because they are less correlated to the rest of the traffic samples and hence, hold more information. The optimal weighting sequence, however, depends on the actual value of $u$ which is obviously not known a priori. Thus, $V_{\tilde{u}_w,N}^*$ serves as a lower bound on the estimation error when weighted sampling is used. Besides, the accuracy of the estimate can be improved in an iterative manner where $\tilde{u}_w$ can be used to calculate an estimate of the optimal weighting sample, then the resulting weighting sequence can be used to improve the estimate of $u$.

\subsection{The Weighted Averaging Estimator under Imperfect Sensing}

In this section, we consider the performance of the weighted averaging estimator considering spectrum sensing errors. The spectrum sensing errors cause the estimator to become biased, akin to the averaging estimator case presented in Section~\ref{sec:estimation_u_err}. Accordingly, the bias can be eliminated by defining the following estimator
\begin{equation}
\tilde u_{w,s}=\frac{1}{1-P_f-Pm}\left[-P_f+\sum_{i=1}^{N}\tilde w_i\tilde z_i\right],
\label{eq:uwe}
\end{equation}
where $\tilde w_i$ is the weight of the estimated traffic sample $\tilde z_i$. Hence, $E[\tilde u_{w,s}]=u$ under the condition $\sum_{i=1}^{N}\tilde w_i=1$. To quantify the MSE in $\tilde u_{w,s}$, we start by evaluating the correlation between the estimated traffic samples, $\tilde R_{i,i+j}$, where
\begin{align}
\tilde R_{i,i+j} & = E[\tilde z_i\tilde z_{i+j}]\nonumber\\ & = \text{Pr}\{\tilde z_i=1, \tilde z_{i+j}=1\}\nonumber\\
& = P_f^2\text{Pr}\{z_i=0, z_{i+j}=0\} + P_f(1-P_m)\nonumber\\&\qquad\times\Big[\text{Pr}\{z_i=0, z_{i+j}=1\} + \text{Pr}\{z_i=1, z_{i+j}=0\}\Big]\nonumber\\& \qquad + (1-P_m)^2\text{Pr}\{z_i=1, z_{i+j}=1\}.
\label{eq:correlation_e}
\end{align}
Recall from Section~\ref{sec:mse_u_weight} that $\text{Pr}\{z_i=1, z_{i+j}=1\} = R_{i,i+j}$. It follows that $\text{Pr}\{z_i=1, z_{i+j}=0\}=u-R_{i,i+j}$, $\text{Pr}\{z_i=0, z_{i+j}=1\}=u-R_{i,i+j}$, and $\text{Pr}\{z_i=0, z_{i+j}=0\} = 1 - 2u + R_{i,i+j}$. Hence, substituting in (\ref{eq:correlation_e}) yields
\begin{align}
\tilde R_{i,i+j} & = R_{i,i+j}(1-P_m-P_f)^2\nonumber\\&\qquad + 2uP_f(1-P_m-P_f)+P_f^2.
\label{eq:correlation:ee}
\end{align}
Since, as in Section~\ref{sec:mse_u_weight}, the estimated samples follow a wide-sense stationary process, then $\tilde R_{i,i+j} = \tilde R[j], \forall j\geq0$. Hence, the variance of the proposed estimator can be expressed as
\begin{align}
V_{\tilde u_{w,s},N} & = E\Big[(\tilde u_{w,s}-u)^2\Big] = E\Big[\tilde u_{w,s}^2\Big]-u^2 \nonumber\\
& = -u^2+\frac{1}{(1-P_f-P_m)^2}\nonumber\\&\times\left(E\Bigg[\left(\sum_{i=1}^N\tilde w_i\tilde z_i\right)^2\Bigg] - 2P_f\sum_{i=1}^N\tilde w_i E\Big[\tilde z_i\Big]+P_f^2\right) \nonumber\\
&=-u^2+\frac{1}{(1-P_f-P_m)^2}\left(\sum_{i=1}^N \tilde w_i^2 E[\tilde z_i^2]\right.\nonumber\\&\left. +2\sum_{j=1}^{N-1}\sum_{i=1}^{N-j}\tilde w_i\tilde w_{i+j} \tilde{R}[j]-2P_f E[\tilde z_i] + P_f^2\right).
\label{eq:V_wee}
\end{align}
Substituting (\ref{eq:correlation:ee}) in (\ref{eq:V_wee}) and replacing $E[\tilde z_i]$ and $E[\tilde z_i^2]$ by $(1-P_f-P_m)u+P_f$ yields
\begin{align}
V_{\tilde u_{w,s},N} & = \left(\frac{uP_m(1-P_m)+(1-u)P_f(1-P_f)}{(1-P_f-P_m)^2}\right)\sum_{i=1}^N\tilde w_i^2\nonumber\\&+ u(1-u)\left(\sum_{i=1}^N\tilde w_i^2+2\sum_{j=1}^{N-1}\sum_{i=1}^{N-j}\tilde w_i\tilde w_{i+j}\right).
\label{eq:V_weee}
\end{align}
In order to derive the optimal weighting sequence that minimizes the MSE in $\tilde{u}_{w,s}$, we apply the orthogonality principle, i.e., $E[(u-\tilde u_{w,s})\tilde z_j]=0, \forall j\in\{1,2,\cdots,N\}$. By solving the normal equations $\tilde{\mathbf{R}} \tilde{\mathbf{w}} = \mathbf{q}_{u\tilde z}$, where $\tilde{\mathbf{R}} \in {\mathbb{R}}^{N\times N}$ is the autocorrelation matrix $\frac{1}{1-P_f-P_m}E[\mathbf{\tilde z}\mathbf{\tilde z}^{T}]$, $\mathbf{\tilde w}$ is the weights vector $[\tilde w_{1}, \tilde w_{2},\cdots,\tilde w_{N}]^{T}$, and $\mathbf{q}_{u\tilde z}$ is the cross correlation vector $E\Big[\left(u+\frac{P_f}{1-P_f-P_m}\right)\mathbf{\tilde z}\Big]$, we derive the normalized optimal weights as $\tilde w_{1}^* = \tilde w_{N}^* = \frac{1}{N(1-\Gamma_c)+2\Gamma_c}$, and $\tilde w_{i}^* = \frac{1-\Gamma_c}{N(1-\Gamma_c)+2\Gamma_c} \text{ } \forall i \notin \{1,N\}$. This is the same result as for the weighted estimator assuming perfect spectrum sensing. Substituting the optimal weights in (\ref{eq:V_weee}) yields
\begin{align}
V_{\tilde u_{w,s},N} & = \frac{u(1-u)(1+\Gamma_c)}{N(1-\Gamma_c)+2\Gamma_c}\nonumber\\&\qquad + \frac{uP_m(1-P_m)+(1-u)P_f(1-P_f)}{(1-P_f-P_m)^2}\nonumber\\&\qquad\times \frac{2+(N-2)(1-\Gamma_c)^2}{(N(1-\Gamma_c)+2\Gamma_c)^2}.
\label{eq:mse_weight_opt_se}
\end{align}
Note that the rightmost part of the right hand side of (\ref{eq:mse_weight_opt_se}) represents the increase in the estimation error attributed to the spectrum sensing errors, while the leftmost part represents the estimation error under perfect sensing as derived in Section~\ref{sec:opt_weight_seq}. It follows that, for a given observation window length, the MSE in $\tilde{u}_{w,s}$ approaches an asymptote, denoted by $V_{\tilde{u}_{w,s},L}^*$, as $N$ increases, due to sample correlation, where $V_{\tilde{u}_{w,s},L}^* = \lim_{N\to\infty} V_{\tilde{u}_{w,s},N}^* = \frac{u(1-u)}{1+\frac{\lambda_f T}{2u}}$.

\paragraph*{Remarks}

The optimal weighting sequence is not affected by the spectrum sensing errors, as the sensing error is assumed to be independent for the different traffic samples. Moreover, as $N$ is increased, the impact of spectrum sensing errors is eliminated where $V_{\tilde{u}_{w,s},L}^* = V_{\tilde{u}_{w},L}^*$. On the other hand, the estimation error caused by the sample correlation serves as a lower bound that can only be decreased by increasing the total observation window length. Furthermore, as in the case of weighted averaging assuming perfect sensing, the optimal weighting sequence depends on the actual value of $u$ which is obviously not known a priori. Thus, $V_{\tilde{u}_{w,s},N}^*$ serves as a lower bound on the estimation error.

\subsection{ML Estimation of $u$ and the CR bound on the estimation error under Perfect Sensing}
\label{sec:ml_u}

In the previous sections, it was shown that the accuracy of the estimators of $u$ that are based on sample stream averaging is limited by sample correlation. Hence, in this section, we propose a more accurate estimator of $u$ based on ML estimation. However, the improved accuracy of ML estimators comes at the expense of an increase in the computational complexity. ML estimation is used to estimate parameters of a statistical model by finding the parameters' values that maximize the probability of the observed samples~\cite{raol_book_2004}. The mean squared estimation error of ML estimators is often quantified analytically using the CR bound. The CR bound quantifies the minimum mean squared estimation error that can be achieved by any unbiased estimator. ML estimators achieve the CR lower bound as the sample size tends to infinity when certain conditions are satisfied~\cite[Ch. 12]{ML_ref}. Accordingly, we present the likelihood function for the estimation of $u$, as well as the corresponding CR bound on the estimation error. The expressions are first presented for the special case when the sensing procedure is assumed to be perfect. Then, the likelihood function is modified to account for the effect of sensing errors on the estimation error. However, the CR bound on the estimation error in the presence of sensing errors cannot be expressed in a simple closed form, hence, the estimation error is presented via simulations in Section~\ref{sec:num_MSE_u_lf_se}. For analytical tractability, uniform sampling is assumed with a constant inter-sample time of $T_c$\,seconds.

We first consider the case where the sensing error can be ignored, i.e., $P_f = P_m = 0$, thus, $\tilde z_n = z_n \forall n$. The likelihood function of the traffic samples given $u$ assuming perfect sensing is derived in a similar manner to~\cite[Sec. 6.1]{kim_tmc_2008} and can be written as
\begin{align}
L(\boldsymbol{z}|u) & = \Pr(\boldsymbol{z}|u) = \Pr(z_1|u) \prod_{i=1}^{N-1} \Pr(z_{i+1} | z_i, u)\nonumber\\& = u^{z_1}\left(1-u\right)^{1-z_1} \prod_{i=1}^{N-1} \textstyle\Pr_{z_{i}z_{i+1}}(T_c|u),
\label{eq:likelihood_func_u}
\end{align}
where the Markovian property has been applied. Expression (\ref{eq:likelihood_func_u}) can be written as
\begin{align}
L(\boldsymbol{z}|u) & = u^{z_1}(1-u)^{1-z_1} \textstyle\Pr_{00}^{n_0}(T_c|u) \textstyle\Pr_{01}^{n_1}(T_c|u)\nonumber\\&\quad\times \textstyle\Pr_{10}^{n_2}(T_c|u) \textstyle\Pr_{11}^{n_3}(T_c|u),
\label{eq:likelihood_func_lambda_f2}
\end{align}
where $n_0$, $n_1$, $n_2$ and $n_3$ denote the number of $(0\!\rightarrow\!0)$, $(0\!\rightarrow\!1)$, $(1\!\rightarrow\!0)$ and $(1\!\rightarrow\!1)$ PU state transitions, respectively, from the total of $N-1$ transitions among $N$ samples. Then, the ML estimator of $u$, denoted by $\tilde{u}_m$, can be found by solving $\partial \log L(\boldsymbol{z}|u)/ \partial u = 0$. The value of $\tilde{u}_m$ cannot be written in a simple closed form and thus, has to be solved numerically. The MSE in $\tilde{u}_m$, denoted by $V_{\tilde{u}_m,N}$, is lower bounded by the CR bound, accordingly, 
\begin{equation}
V_{\tilde{u}_m,N} \geq I_m^{-1}\left(u,N\right),
\label{eq;u_cr1}
\end{equation}
where $I_m\left(u,N\right)$ is the Fisher information for $N$ collected samples. Specifically, the Fisher information is defined as $I_m(u,N) = E\left[\left(\partial\log L(\boldsymbol{z}|u)/\partial u\right)^2\right]$.
\begin{theorem}
The lower bound on the MSE in $\tilde{u}_m$ is given as
\begin{align}
I_m^{-1}\left(u,N\right) &= \frac{\left(1-\Gamma_c\right)\left(\Gamma_c+u-\Gamma_c u\right)\left(\Gamma_c u-u+1\right)}{M_1 + M_2 + M_3 + M_4 + M_5}\nonumber\\&\qquad\times u^3 \left(1-u\right),
\label{eq;nb_4_u}
\end{align}
where $M_1=\Gamma_c^2 \lambda_f T_c(N-1)(1-u)[\lambda_f T_c(1-u)(1+\Gamma_c )-2u(1-2u)(1-\Gamma_c )]$, $M_2=\Gamma_c^3 u^2 [u(u-1)(3N-2)+(N-1)]$, $M_3 = -\Gamma_c^2 u^2 [u(u-1)(7N-4)+(2N-1)]$, $M_4 = \Gamma_c u^2 [N(5u^2-5u+1)+2u(1-u)]$, and $M_5 = Nu^3 (1-u)$.

\begin{IEEEproof}
See Appendix~\ref{sec:math_ind_u}.
\end{IEEEproof}
\end{theorem}

For a fixed observation window, the CR bound for the variance in $\tilde{u}_m$ approaches an asymptote as $N$ increases. This is caused by the correlation between the samples. The lower bound on the CR bound can be derived as follows
\begin{equation}
\lim_{N\to\infty} I_m^{-1}\left(u,N\right) = \frac{u(1-u)}{1+\frac{\lambda_f T}{u}}.
\label{eq;nb_5_u}
\end{equation}
\paragraph*{Remarks}

Comparing (\ref{eq;nb_5_u}) to (\ref{eq;MSE_w_inf}), as $N$ approaches infinity, the MSE in estimating $u$ using ML estimation with an observation window length of $T$ is the same as that when weighted averaging is used with an observation window length of $2T$. This is a very useful property for delay-constrained applications as the estimation accuracy can be achieved in half the time. The fact that ML estimation requires only half the time window length is attributed to the fact that the proposed ML estimator assumes knowledge of either $\lambda_f$ or $\lambda_n$, hence, half the information needed to estimate the traffic parameters is assumed to be known a priori. The assumption of knowing $\lambda_f$ or $\lambda_n$ beforehand is application specific (the average PU off- or on- time may be known to the SU in some network scenarios), where on the other hand, in case both $\lambda_f$ and $\lambda_n$ are unknown, joint ML estimation should be used.

\subsection{ML Estimation of $u$ under Imperfect Sensing}
\label{sec:ml_u_se}

In the presence of sensing errors, any PU traffic samples vector $\boldsymbol{z}$ can result in an estimated PU traffic samples vector $\boldsymbol{\tilde z}$ with a non-zero probability. Hence, the likelihood function presented in (\ref{eq:likelihood_func_lambda_f2}) is modified to
\begin{equation}
L(\boldsymbol{\tilde z}|u) = \sum_{n=1}^{2^N} \Pr(\mathcal{Z}_n|u) S(\boldsymbol{\tilde z}|\mathcal{Z}_n),
\label{eq;joint_error_ss_1}
\end{equation}
where $\Pr(\mathcal{Z}_n|u)$ is the probability of occurrence of the PU traffic samples vector $\mathcal{Z}_n$ and equals the right hand side of (\ref{eq:likelihood_func_lambda_f2}), and $S(\boldsymbol{\tilde z}|\mathcal{Z}_n)$ is the probability of estimating the PU traffic samples vector as $\boldsymbol{\tilde z}$, when the actual PU traffic samples vector equals $\mathcal{Z}_n$. $S(\boldsymbol{\tilde z}|\mathcal{Z}_n)$ can be written as
\begin{align}
S(\boldsymbol{\tilde z}|\mathcal{Z}_n) & = P_{f}^{m_{0,n,\boldsymbol{\tilde z}}} (1 - P_{f})^{m_{1,n,\boldsymbol{\tilde z}}} \nonumber\\&\quad\times P_{m}^{m_{2,n,\boldsymbol{\tilde z}}} (1 - P_{m})^{m_{3,n,\boldsymbol{\tilde z}}},
\label{eq;joint_error_ss_2}
\end{align}
where $m_{0,n,\boldsymbol{\tilde z}}$, $m_{1,n,\boldsymbol{\tilde z}}$, $m_{2,n,\boldsymbol{\tilde z}}$, and $m_{3,n,\boldsymbol{\tilde z}}$ are the numbers of false alarms, no false alarms, mis-detections, and no mis-detections, respectively, that yield the estimated PU traffic samples vector $\boldsymbol{\tilde z}$ given the PU traffic samples vector $\mathcal{Z}_n$. The ML estimator of $u$ can be modified to account for sensing errors and can be calculated by solving for the value of $u$ that maximizes the modified likelihood function given in (\ref{eq;joint_error_ss_1}). The ML estimator as well as the corresponding mean squared estimation error cannot be expressed in a simple closed form and, hence, have to be calculated numerically as shown in Section~\ref{sec:numerical_results}.

\section{Estimation of the Primary User Departure Rate $\lambda_{f}$ and Arrival Rate $\lambda_{n}$}
\label{sec:lambda_f_estimation}

\subsection{Estimation of $\lambda_{f}$ under Perfect Sensing}
\label{sec:lambda_f_estimation_ML}

Following~\cite[Sec. 6.1]{kim_tmc_2008} the estimation method adopted here is based on ML estimation. Uniform sampling is assumed for mathematical tractability. The likelihood function of the traffic samples given $\lambda_f$ is the same as that in (\ref{eq:likelihood_func_lambda_f2}) but with replacing the condition on $u$ by a condition on $\lambda_f$. Then, the ML estimator of $\lambda_f$, denoted by $\tilde\lambda_f$, can be found by solving $\partial \log L(\boldsymbol{z}|\lambda_{f})/ \partial \lambda_f = 0$ as in~\cite[Sec. 6.1]{kim_tmc_2008} yielding $\tilde\lambda_f=-\left(u/T_c\right)\log\left[\left(-B+\sqrt{B^2-4AC}\right)/\left(2A\right)\right]$, where $A=(u-u^2)(N-1)$, $B=-2A+N-1-(1-u)n_0-un_3$, and $C=A-un_0-(1-u)n_3$.

\subsection{The CR bound on the MSE in $\tilde\lambda_{f}$}
\label{sec:lambda_f_estimation_error}

The MSE in $\tilde\lambda_{f}$, denoted by $V_{\tilde{\lambda}_f,N}$, is expressed using the CR bound as in the case of the ML estimation of $u$. Hence, 
\begin{equation}
V_{\tilde{\lambda}_f,N} \geq I_m^{-1}\left(\lambda_f,N\right),
\label{eq;lf_cr1}
\end{equation}
where $I_m\left(\lambda_f,N\right) = E\left[\left(\partial\log L(\boldsymbol{z}|\lambda_{f})/\partial\lambda_{f}\right)^2\right]$. Accordingly, we obtain the following theorem.
\begin{theorem}
The lower bound on the MSE in $\tilde\lambda_f$ is given as
\begin{equation}
I_m^{-1}\left(\lambda_f,N\right) = \frac{u(1-\Gamma_c)[\Gamma_c+u(1-\Gamma_c)^2(1-u)]}{\left(\Gamma_c T_c\right)^2 (1-u)(1+\Gamma_c)(N-1)}.
\label{eq;nb_4}
\end{equation}
\begin{IEEEproof}
See Appendix~\ref{sec:math_ind_lf}.
\end{IEEEproof}
\end{theorem}
As for the case of estimating $u$, the CR bound for the variance in $\tilde\lambda_f$ for a fixed observation window approaches an asymptote as $N$ increases due to the sample correlation. The lower limit on the CR bound can be derived by taking the limit of (\ref{eq;nb_4}) as $N$ approaches infinity, yielding
\begin{equation}
\lim_{N\to\infty} I_m^{-1}\left(\lambda_f,N\right) = \frac{\lambda_f}{2T\left(1-u\right)}.
\label{eq;nb_5}
\end{equation}

\subsection{Estimation of $\lambda_{n}$ under Perfect Sensing}
\label{sec:lambda_n_estimation_ML}

The PU arrival rate, $\lambda_{n}$, can be estimated in a similar manner to $\lambda_{f}$ using ML estimation. The ML estimator of $\lambda_{n}$ can be written as $\tilde\lambda_{n} = \left(1-u\right)\tilde\lambda_{f}/u$ where the derivation follows that of $\tilde\lambda_{f}$ and is omitted for brevity. It follows that $I_m\left(\lambda_n,N\right) = u^2 I_m\left(\lambda_f,N\right)/\left(1-u\right)^2$. Moreover, for a fixed observation window, the CR bound for the MSE in $\tilde\lambda_{n}$ approaches $\lambda_n/\left(2Tu\right)$ as $N$ increases.

\paragraph*{Remarks}
\label{sec:lambda_f_estimation_ML_Remarks}

The expressions for $\tilde\lambda_{f}$ and $\tilde\lambda_{n}$ are functions of $u$, thus, ML estimation can be used in applications where a priori knowledge of $u$ can be assumed. Moreover, the expressions for the MSE in $\tilde\lambda_{f}$ and $\tilde\lambda_{n}$ are functions of the actual values of $\lambda_{f}$ and $\lambda_{n}$, which are not known a priori. Hence, the MSE expressions can be used to provide benchmarks for the ML estimation error. Furthermore, the MSE expressions can be used to calculate the worst case error in $\tilde\lambda_{f}$ and $\tilde\lambda_{n}$, and to show the dependence of the estimation error on $u$, $\lambda_f$, $\lambda_n$, the number of samples, and the observation window length.

\subsection{ML Estimation of $\lambda_{f}$ and $\lambda_{n}$ under Imperfect Sensing}
\label{sec:ml_lambda_se}

The likelihood function of the traffic samples given $\lambda_f$ or $\lambda_n$ under imperfect sensing can be expressed as in (\ref{eq;joint_error_ss_1}) with replacing the condition on $u$ by a condition on $\lambda_f$ and $\lambda_n$, respectively. The ML estimators for $\lambda_f$ and $\lambda_n$ as well as the corresponding mean squared estimation errors cannot be expressed in simple closed forms and consequently, have to be calculated numerically as shown in Section~\ref{sec:numerical_results}.

\section{Algorithms for the Blind Estimation of $u$, $\lambda_f$, and $\lambda_n$}
\label{sec:algorithm}

In this section, we present algorithms that blindly estimate $u$, $\lambda_f$ and $\lambda_n$ based on adaptive sampling, using the analytical expressions obtained thus far. The assumptions that are necessary for the operation of the algorithms are that the off- and on-times of the PU are exponentially distributed. Besides, perfect spectrum sensing is assumed, noting that the algorithms can be updated to account for the effect of spectrum sensing imperfections. Two algorithms are presented: Algorithm $\bf{I}$ blindly estimates $u$ assuming perfect knowledge of $\lambda_f$ and no a priori knowledge of $\lambda_n$ (see the verbal description in Section~\ref{sec:algorithm_I} and its summary in Algorithm~\ref{alg:estimation_wesam_1}), and Algorithm $\bf{II}$ blindly estimates $u$, $\lambda_f$ and $\lambda_n$ (see the verbal description in Section~\ref{sec:algorithm_II} and its summary in Algorithm~\ref{alg:estimation_wesam_2}). 

\subsection{Algorithm $\bf{I}$: Blind Estimation of $u$ with Known $\lambda_f$ (or $\lambda_n$)}
\label{sec:algorithm_I}

Algorithm $\bf{I}$ is applicable in scenarios where there is a priori knowledge of $\lambda_f$ whereas $u$ and $\lambda_n$ are unknown. Note that, the algorithm can be modified to estimate $u$ under the assumption of perfect knowledge of $\lambda_n$ with no a priori knowledge of $\lambda_f$. A practical example for the latter case would be if the average on-time of the PU is known (for example, the packet length of the PU follows a certain pattern or is fixed) while the rate at which the PU accesses the channel is unknown. The algorithm estimates $u$ using the averaging method as described in (\ref{eq;u_est}), and the error in $\tilde{u}_a$ is estimated using (\ref{eq;nb_1}).

The algorithm operates as follows. The traffic is sampled with an arbitrary initial inter-sample time, $T_0$, until the sampled traffic state toggles. Then, traffic is sampled for an arbitrary initial number of samples, $N_0$, with the inter-sample time $T_0$. Note than increasing $T_0$ and $N_0$ increases the accuracy of the initial estimate, but, on the other hand, increases the estimation delay. For the remainder of the algorithm, the inter-sample time is determined while taking the correlation between the traffic samples into consideration. Corollary~\ref{cor_3} provides an expression for the expected decrease in the MSE in $\tilde{u}_a$ for each additional sample. The maximum decrease in the MSE in $\tilde{u}_a$ is given by
\begin{align}
D_{\tilde{u}_a,N+1}^{\max} & = \lim_{T_N\to\infty} V_{\tilde{u}_a,N} - V_{\tilde{u}_a,N+1}\nonumber\\& = \frac{V_{\tilde{u}_a,N}(2N+1)-u(1-u)}{(N+1)^2}.
\end{align}
Accordingly, the parameter $D_{\tilde{u}_a,N+1}$ can be expressed as $D_{\tilde{u}_a,N+1} = D_{\tilde{u}_a,N+1}^{\max} - D_0 \Gamma_N$, where $D_0 = \frac{2u(1-u)}{(N+1)^2}\sum_{i=0}^{N-1} \prod_{k=N-i}^{N-1} \Gamma_k$. Parameter $D_{\tilde{u}_a,N+1}^{\max}$ is independent of $T_N$, thus, varying $T_N$ can only affect the $D_0 \Gamma_N$ term in $D_{\tilde{u}_a,N+1}$. Setting $T_N$ to $\frac{u\alpha}{\lambda_f}$ is equivalent to multiplying $D_0$ by a factor of $e^{-\alpha}$. The parameter $\alpha$ is chosen so that a compromise between the reduction in the estimation error per sample and the delay in taking the new sample is reached. For lower values of $\alpha$, the expected decrease in the MSE in $\tilde{u}_a$ per sample is lower, while at the same time, the average delay in taking the new sample is lower.

There are three different conditions for terminating the algorithm that are used depending on the application. The algorithm may terminate when (i) a predetermined number of samples are taken (energy-constrained applications), or (ii) after a certain observation window length is reached (delay-constrained applications), or (iii) when a target expected estimation error is reached. For an energy-constrained application where the sensing energy budget, and hence the total number of samples that are to be taken, is limited, the first termination condition is used where the total number of samples equal $N_{th}$. On the other hand, for a delay-constrained application where the total observation window is bounded by a time threshold, $T_{th}$, the second termination condition is applied. Finally, the third termination condition is used if the application involves estimating $u$ with a target average estimation error $V_{th}$. To ensure that the target average error is always reached, the expected error is calculated using (\ref{eq;nb_1}) for $u\in[0,1]$, using the operating values of $N$, $\lambda_f$, and $\mathcal{T}$. The algorithm is terminated if the worst case expected estimation error, calculated over all possible values of $u$, is less than $V_{th}$. Note that the value of $u$ that yields the highest estimation error cannot be known a priori as described in Section~\ref{sec:num_MSE_u}.The proposed algorithm is summarized in Algorithm~\ref{alg:estimation_wesam_1}.

\begin{algorithm}[t]
\centering
\caption{Algorithm $\bf{I}$: Blind Estimation of $u$ with known $\lambda_f$ (or $\lambda_n$)}
\label{alg:estimation_wesam_1}
\begin{algorithmic}[1]
\REQUIRE $T_0$, $N_0$, $\alpha$
\ENSURE $T < T_{th}$ or $N < N_{th}$ or max\{$V_{\tilde{u}_a,N}\}<V_{th}$ (user specified)
\WHILE {$\tilde{u}_a = 0$ or $\tilde{u}_a = 1$}
	\STATE {take a next sample after $T_0$\,s}
	\STATE {calculate $\tilde{u}_a$ using (\ref{eq;u_est})}
	\IF{$T\geq T_{th}$ or $N\geq N_{th}$ (if target $T_{th}$ or $N_{th}$ is required)}
	\STATE Terminate
	\ENDIF
\ENDWHILE
\WHILE {$N<N_0$}
	\STATE {take a next sample after $T_0$\,s}
	\STATE {calculate $\tilde{u}_a$ using (\ref{eq;u_est})}
	\IF{$T\geq T_{th}$ or $N\geq N_{th}$ (if target $T_{th}$ or $N_{th}$ is required)}
	\STATE Terminate
	\ENDIF
\ENDWHILE
\WHILE {$T<T_{th}$ or $N<N_{th}$ or max\{$V_{\tilde{u}_a,N}\}>V_{th}$ (user specified)}
	\STATE {calculate $T_{N} = \frac{\tilde{u}_a}{\lambda_f}\alpha$}
	\STATE {Take a next sample after $T_{N}$\,s}
	\STATE {calculate $\tilde{u}_a$ using (\ref{eq;u_est})}
	\STATE {calculate $V_{\tilde{u}_a,N}$ for all $u\in[0,1]$ using (\ref{eq;nb_1}) (if target $V_{th}$ is required)}
\ENDWHILE
\end{algorithmic}
\end{algorithm}

\subsection{Algorithm $\bf{II}$: Blind Estimation of $u$, $\lambda_f$ and $\lambda_n$}
\label{sec:algorithm_II}

Algorithm $\bf{II}$ is directed for scenarios where there is no a priori knowledge of $u$, $\lambda_f$, and $\lambda_n$. The duty cycle is estimated using the averaging method as described in (\ref{eq;u_est}). Parameters $\lambda_f$ and $\lambda_n$ are estimated using ML estimation as described in Section~\ref{sec:lambda_f_estimation}, hence, unlike Algorithm $\bf{I}$, the inter-sample time is kept constant at $T_0$. The algorithm terminates when a target average estimation error in $u$, denoted by $V_{u,th}$, and a target average estimation error in $\lambda_f$ or $\lambda_n$, denoted by $V_{\lambda,th}$, are reached. To ensure that the targeted estimation errors are met, the estimation errors are calculated at all algorithm iterations for the full range of $u$, and $\lambda_f$ or $\lambda_n$. Moreover, it follows from Section~\ref{sec:lambda_f_estimation} that the asymptotic lower bound on the error in $\lambda_f$ is greater than that in $\lambda_n$ for $u>\frac{1}{2}$ and approaches infinity as $u$ tends to $1$, while the asymptotic lower bound on the error in $\lambda_n$ is greater than that in $\lambda_f$ for $u<\frac{1}{2}$ and approaches infinity as $u$ tends to $0$. Thus, the algorithm is designed to terminate when $\tilde{u}_a$ and $\tilde{\lambda}_f$ reach the target average estimation error if $\tilde{u}_a<\frac{1}{2}$, and when $\tilde{u}_a$ and $\tilde{\lambda}_n$ reach the target average estimation error if $\tilde{u}_a>\frac{1}{2}$. Besides, The lower bound on the estimation error for $\lambda_f$ and $\lambda_n$ is proportional to $\lambda_f$ and $\lambda_n$ as shown in Section~\ref{sec:lambda_f_estimation}. Accordingly, the error in $\lambda_f$ and $\lambda_n$ cannot be guaranteed to meet specific targets unless $\lambda_f$ and $\lambda_n$ are upper bounded. Moreover, we assume that $\lambda_f$ and $\lambda_n$ are greater than zero, as otherwise, the PU would be either always on or always off. Thus, in this section we assume that $\lambda_f, \lambda_n \in [\lambda_{\min}, \lambda_{\max}]$. Furthermore, since the error in estimating $u$ increases with decreasing $\lambda_f$ as presented in (\ref{eq;nb_1}), the worst case estimation error in $u$ is calculated while substituting $\lambda_f=\lambda_{\min}$. On the other hand, as the estimation error in $\lambda_f$ and $\lambda_n$ increases monotonically with $\lambda_f$ and $\lambda_n$, respectively, the worst case estimation error in $\lambda_f$ and $\lambda_n$ is calculated while substituting $\lambda_f=\lambda_n=\lambda_{\max}$. The proposed algorithm is summarized in Algorithm~\ref{alg:estimation_wesam_2}.

\begin{algorithm}[t]
\centering
\caption{Algorithm $\bf{II}$: Blind Estimation of $u$, $\lambda_f$, and $\lambda_n$}
\label{alg:estimation_wesam_2}
\begin{algorithmic}[1]
\REQUIRE $T_0$
\ENSURE max\{$V_{\tilde{u}_a,N}\}<V_{u,th}$ and max\{$V_{\tilde{\lambda}_f,N}\}<V_{\lambda,th}$ for $\tilde{u}_a<0.5$, and max\{$V_{\tilde{u}_a,N}\}<V_{u,th}$ and max\{$V_{\tilde{\lambda}_n,N}\} \text{ (MSE in $\tilde{\lambda}_n$) } < V_{\lambda,th}$ for $\tilde{u}_a>0.5$
\WHILE {$\tilde{u}_a = 0$ or $\tilde{u}_a = 1$}
	\STATE {take a next sample after $T_0$\,s}
	\STATE {calculate $\tilde{u}_a$ using (\ref{eq;u_est})}
\ENDWHILE
\WHILE {\{$\tilde{u}_a<0.5$, and max\{$V_{\tilde{u}_a,N}\}>V_{u,th}$ or max\{$V_{\tilde{\lambda}_f,N}\}>V_{\lambda,th}$\} or \{$\tilde{u}_a>0.5$, and max\{$V_{\tilde{u}_a,N}\}>V_{u,th}$ or max\{$V_{\tilde{\lambda}_n,N}\} >V_{\lambda,th}$\}}
	\STATE {take a next sample after $T_0$\,s}
	\STATE {calculate $\tilde{u}_a$ using (\ref{eq;u_est})}
	\STATE {calculate $V_{\tilde{u}_a,N}$ for all $u\in[0,1]$ and for $\lambda_f=\lambda_{\min}$ using (\ref{eq;nb_1})}
	\STATE {calculate $\tilde{\lambda}_f$ and $\tilde{\lambda}_n$ following Section~\ref{sec:lambda_f_estimation} using $\tilde{u}_a$, $N$, $n_0$, and $n_3$}
	\STATE {calculate $V_{\tilde{\lambda}_f,N}$ and $V_{\tilde{\lambda}_n,N}$ for all $u\in[0,1]$ and for $\lambda_f=\lambda_n=\lambda_{\max}$ following Section~\ref{sec:lambda_f_estimation}}	
\ENDWHILE
\end{algorithmic}
\end{algorithm}

\section{Numerical Results}
\label{sec:numerical_results}

We present numerical results, confirming the theory provided in Section~\ref{sec:squared_error}, that compare the accuracy of the different estimation methods of $u$. Moreover, we show the accuracy and the asymptotic behavior of the ML estimation of $\lambda_f$, based on the theory obtained in Section~\ref{sec:lambda_f_estimation}. Furthermore, we present results for the estimation error in $u$ and $\lambda_f$ under spectrum sensing imperfections. Note that the performance of the estimation of $\lambda_n$ is similar to that of $\lambda_f$, and thus is omitted to eliminate redundancy. Next, we present results showing the performance of the proposed blind estimation algorithms. Besides, we validate the correctness of the developed mathematical expressions through comparison with simulation results developed in Matlab version 7.10.0.499. Note that, the root mean squared (RMS) error is used as a metric for quantifying the estimation accuracy instead of the MSE for convenience. The RMS error is simply calculated as the square root of the MSE. 

In the numerical evaluation, as an example, typical values for the PU traffic parameters are used following the results in~\cite{liang_tmc_2011,wellens_phycom_2009}, which are representative sources of information of the temporal utilization of radio resources in popular radio access systems. In specific, we focus on GSM 1800 downlink traffic, as it is the only service common to both studies, with detailed parameters found in~\cite[Tab. VIII]{liang_tmc_2011} and~\cite[Tab. 3]{wellens_phycom_2009}. We therefore assume that the PU duty cycle, $u$, is in the range of [0.3,0.6] (0.30 in~\cite{liang_tmc_2011} and 0.62 in~\cite{wellens_phycom_2009}), whereas the PU departure rate, $\lambda_f$, is in the range\footnote{As the measurements in~\cite{wellens_phycom_2009} were performed in the discrete domain, we directly converted the parameters of all geometric distributions given in~\cite[Tab. 3]{wellens_phycom_2009} into exponential distribution parameters, where the minimum and maximum reported PU departure rate is 0.48\,s$^{-1}$ and 0.9\,s$^{-1}$, respectively. Note that the arrival and departure rates are not reported explicitly in~\cite{liang_tmc_2011}.} of [0.4,0.9]\,s$^{-1}$.

\subsection{The RMS Error in Estimating $u$}
\label{sec:num_MSE_u}

\subsubsection{The Variation of the RMS Error with the Number of Samples}

\begin{figure}
\centering
\subfigure[$u=0.3$, $\lambda_f \in\{0.4,0.9\}$\,s$^{-1}$]{\includegraphics[width=\columnwidth]{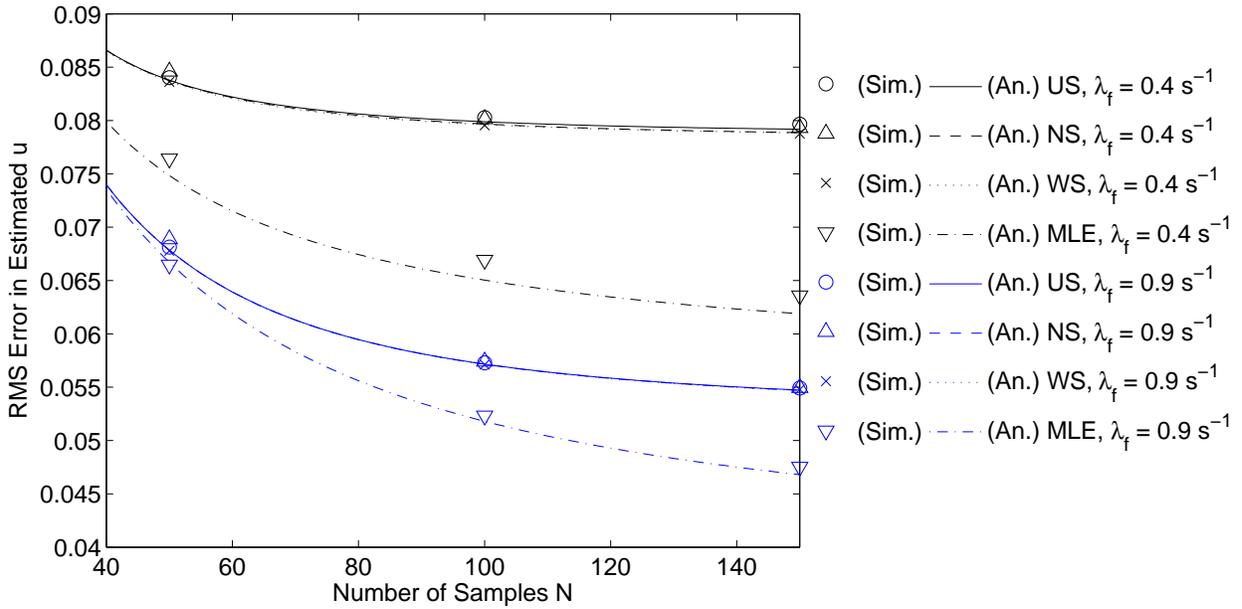}\label{fig:mse_u_N1}}\\
\subfigure[$u=0.6$, $\lambda_f \in\{0.4,0.9\}$\,s$^{-1}$]{\includegraphics[width=\columnwidth]{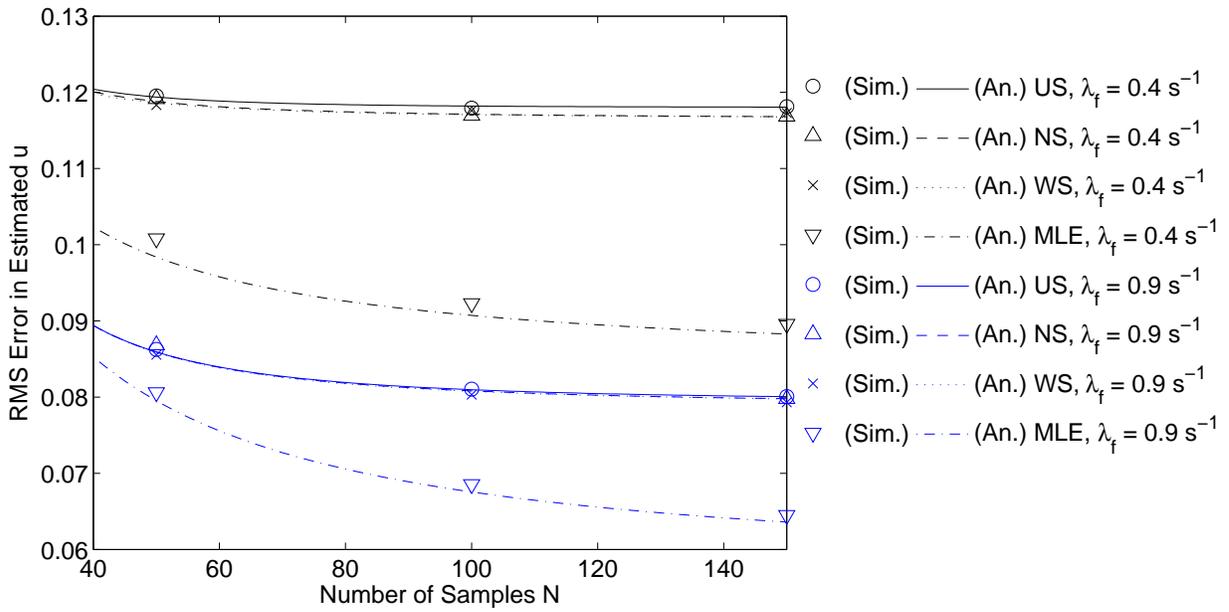}\label{fig:mse_u_N2}}
\caption{RMS error in the estimate of $u$ as a function of the number of samples $N$ for $T=50$\,s and different values of $u$. The RMS error is plotted for the four estimation methods; US: Averaging with uniform sampling, NS: Averaging with non-uniform sampling, WS: Weighted averaging, and MLE: Maximum likelihood estimation. Simulation results (Sim.) are plotted to verify the mathematical model (An.).}
\label{fig:rms_n}
\end{figure}

The relationships between the total number of samples, $N$, and the RMS error in the estimate of $u$ are plotted in Fig.~\ref{fig:rms_n} for (i) the averaging estimator with non-uniform sampling using the optimal inter-sample time sequence, (ii) the averaging estimator with uniform sampling, (iii) the weighted averaging estimator using the optimal weighting sequence, and (iv) the ML estimator. The RMS error in estimating $u$, denoted by $R_{\tilde{u},N}$, is plotted for an observation window of $T=50$\,s where $N$ is increased from 40 to 150 samples, which represents a typical size of sample sets in traffic sampling. The variation of $R_{\tilde{u},N}$ with $N$ for $u=0.3$ and $u = 0.6$ is shown in Fig.~\ref{fig:mse_u_N1} and Fig.~\ref{fig:mse_u_N2}, respectively. The PU departure rate, $\lambda_f$ is set to 0.4\,s$^{-1}$ and 0.9\,s$^{-1}$.

The results show that ML estimation outperforms all averaging based estimation techniques where the resulting RMS error can be reduced by up to 24\%. This is because the proposed ML estimator assumes a priori knowledge of $\lambda_f$. Moreover, the optimized averaging estimator with non-uniform sampling and the optimized weighted averaging estimator yield almost the same estimation error, with a narrow margin below the averaging estimator with uniform sampling. Besides, for the same $u$, higher $\lambda_f$ yields lower estimation error due to the reduced sample correlation. Furthermore, the figures emphasize the fact that the estimation error reaches an asymptotic value as $N$ is increased. Finally, all results are verified via Matlab simulations where the simulation results match the theoretical expressions except for ML estimation where the simulation-based error is higher than the theoretical expression. This is because the CR bound provides a lower bound on the error that is attained asymptotically as $N$ increases.

\subsubsection{The Asymptotic RMS Error}

\begin{figure}
\centering
\includegraphics[width=\columnwidth]{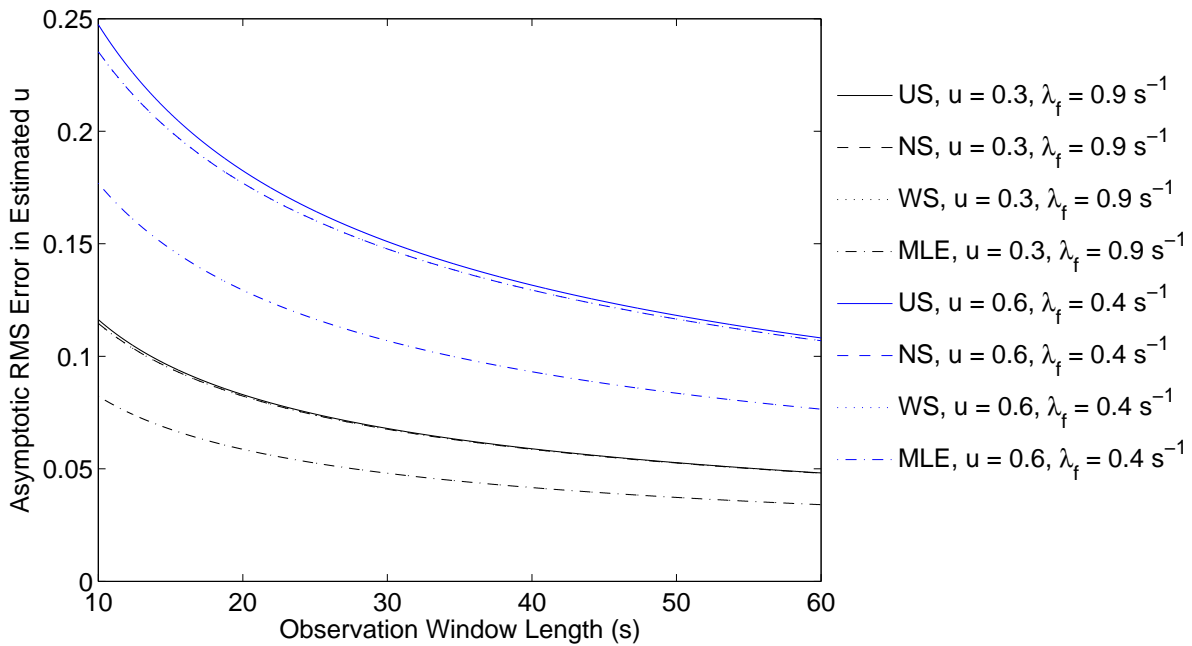}
\caption{Asymptotic RMS error in the estimate of $u$, $R_{\tilde{u},\infty}$, as a function of the observation window length, $T$. $R_{\tilde{u},\infty}$ is plotted for the four estimation methods; US: Averaging with uniform sampling, NS: Averaging with non-uniform sampling, WS: Weighted averaging, and MLE: Maximum likelihood estimation. The traffic parameters used: $\{u = 0.3, \lambda_f = 0.9$\,s$^{-1}\}$, and $\{u = 0.6, \lambda_f = 0.4$\,s$^{-1}\}$.}
\label{fig:asym_T}
\end{figure}

The RMS error in the estimate of $u$ as $N$ tends to infinity reaches an asymptote as shown in Section~\ref{sec:squared_error}. Fig.~\ref{fig:asym_T} shows the relationship between the asymptotic RMS error in the estimate of $u$, denoted by $R_{\tilde{u},\infty}$, and the observation window length, $T$, for the different estimation techniques. Two different sets of PU traffic parameters are used; $u = 0.3$ and $\lambda_f = 0.9$\,s$^{-1}$, and $u = 0.6$ and $\lambda_f = 0.4$\,s$^{-1}$. $R_{\tilde{u},\infty}$ for the averaging estimator with non-uniform sampling is calculated numerically since a closed form expression is not available. The results show that $R_{\tilde{u},\infty}$ for ML estimation is lower than that for the other estimation techniques, and the performance of the weighted averaging estimator and the averaging estimator with non-uniform sampling is almost identical and surpasses that of the averaging estimator with uniform sampling. Moreover, as proven in Section~\ref{sec:ml_u}, $R_{\tilde{u},\infty}$ for ML estimation is identical to that for weighted averaging estimation if the observation window length, $T$, is doubled. 

\subsubsection{The Variation of the RMS Error with the Duty Cycle}

\begin{figure}
\centering
\includegraphics[width=\columnwidth]{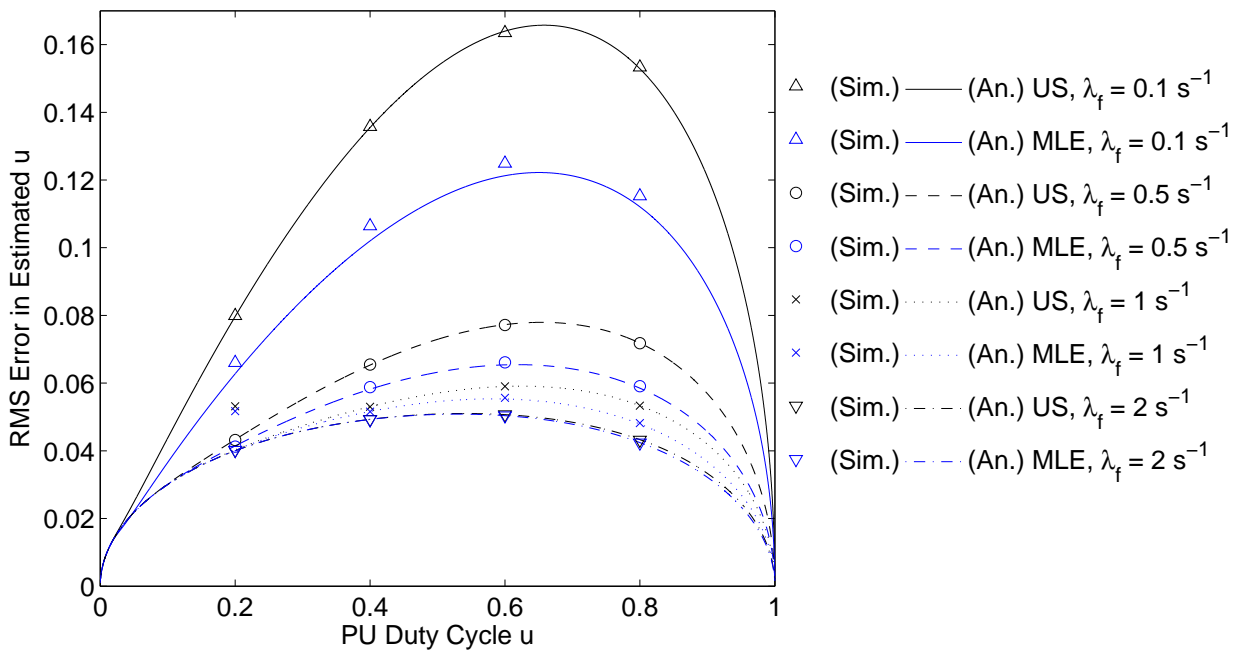}
\caption{RMS error in the estimate of $u$ as a function of $u$ for $N =100$ and $T=100$\,s. The RMS error is plotted for averaging with uniform sampling (US) and the CR bound is plotted for maximum likelihood estimation (MLE). The traffic parameters used: $\lambda_f\in\{0.1,0.5,1,2\}$\,s$^{-1}$. Simulation results (Sim.) are plotted to verify the mathematical model (An.).}
\label{fig:mse_u_u}
\end{figure}

The relationship between $R_{\tilde{u},N}$ and $u$, is presented in Fig.~\ref{fig:mse_u_u}. The plot compares the error for the averaging estimator with uniform sampling with that of the ML estimator. For this setup, $u$ is increased from 0 to 1 while $T$ and $N$ are kept constant at 100 seconds and 100 samples, respectively. The error is presented for different values of $\lambda_f$ representing traffic with different levels of correlation. ML estimation achieves a more accurate estimate than averaging-based estimation, yet the gap in performance decreases with higher $\lambda_f$. The value of $u$ which results in the highest estimation error is greater than 0.5 and approaches $u=0.5$ with increasing $\lambda_f$. The skew in the figure is attributed to the error added by sample correlation, which increases with $u$ for the same $\lambda_f$. For higher $\lambda_f$, the effect of sample correlation decreases and the mean squared estimation error approaches $\frac{u(1-u)}{N}$, as explained in Section~\ref{sec:mse_u_non_uni}, which is symmetric in $u$ and is maximized at $u=0.5$. Again, the results are verified via simulations where the simulation results match the theoretical expressions except for ML estimation as the CR bound provides a theoretical lower bound on the error.

\paragraph*{Key Message}

ML estimation is recommended for estimating $u$ when a priori knowledge of $\lambda_f$ or $\lambda_n$ is available. Moreover, optimized averaging under non-uniform sampling and optimized weighted averaging yield almost the same estimation error, and result in a lower estimation error than averaging under uniform sampling.

\subsection{The RMS Error in $\tilde{\lambda}_f$ under Uniform Sampling}
\label{sec:num_MSE_lf}

\begin{figure}
\centering
\subfigure[CR bound on the RMS error in $\tilde{\lambda}_f$ as a function of $N$ for $T=50$\,s.]{\includegraphics[width=0.49\columnwidth]{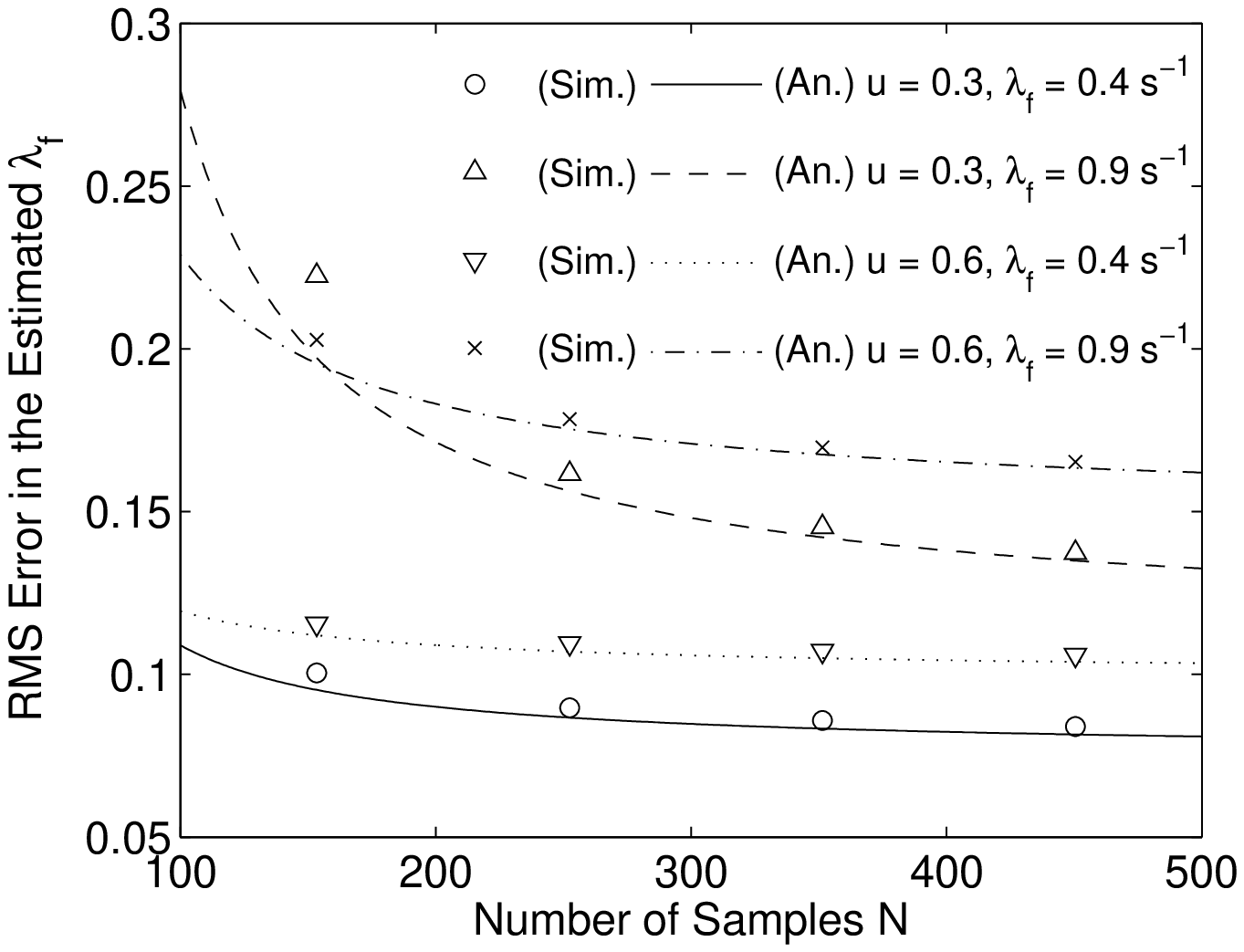}\label{fig:SD_lf_vs_N}}
\subfigure[Asymptotic CR bound on the RMS error in $\tilde{\lambda}_f$ as a function of $T$.]{\includegraphics[width=0.49\columnwidth]{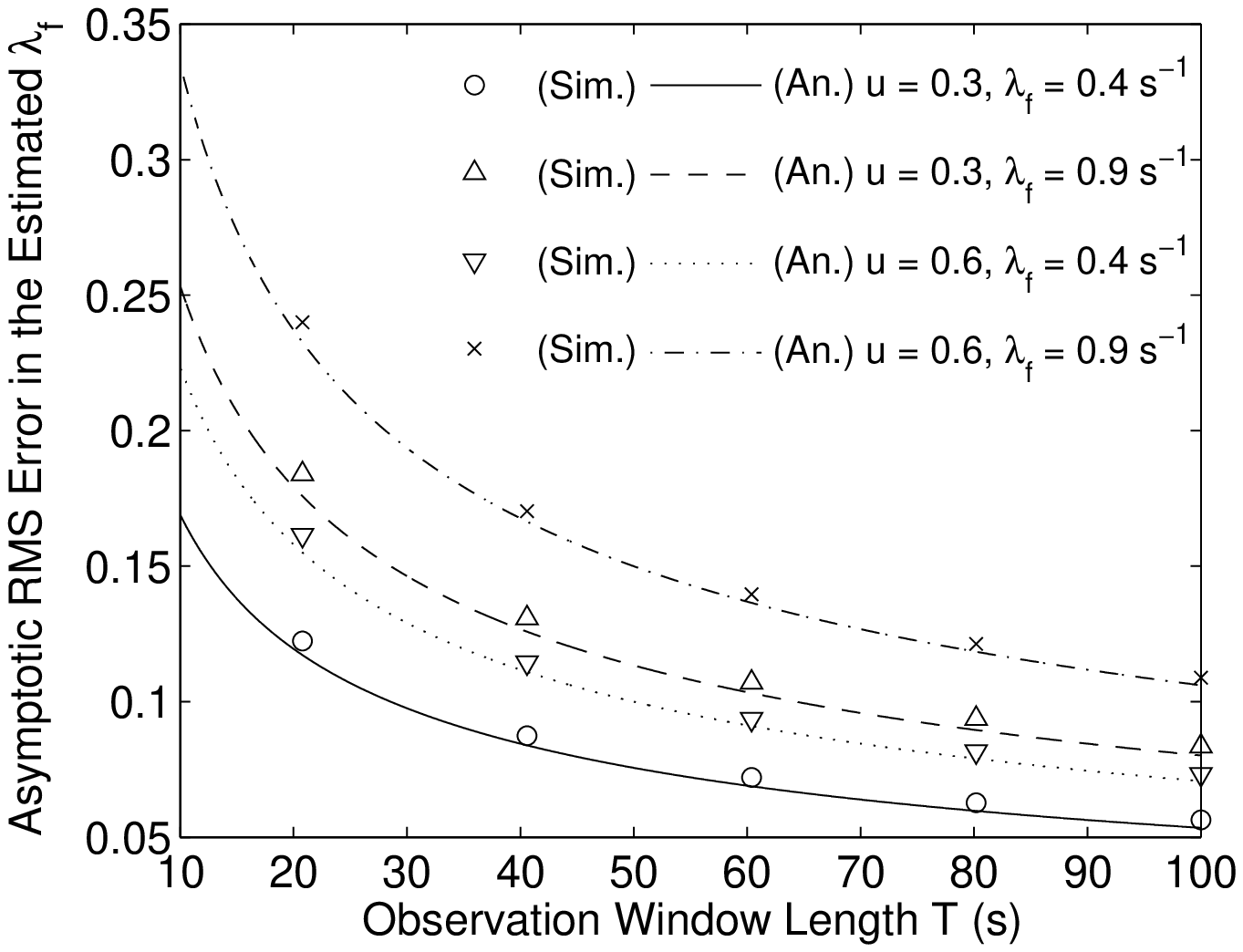}\label{fig:asym_lf}}
\caption{CR bound~\ref{fig:SD_lf_vs_N} and asymptotic CR bound~\ref{fig:asym_lf} on the error in $\tilde{\lambda}_f$. The used traffic parameters: $\{u = 0.3, \lambda_f = 0.4$\,s$^{-1}\}$, $\{u = 0.3, \lambda_f = 0.9$\,s$^{-1}\}$, $\{u = 0.6, \lambda_f = 0.4$\,s$^{-1}\}$, and $\{u = 0.6, \lambda_f = 0.9$\,s$^{-1}\}$. Simulation results (Sim.) are plotted to verify the mathematical model (An.).}
\end{figure}

The PU departure rate, $\lambda_f$, is estimated using ML estimation and hence, the estimate accuracy is lower bounded by the CR bound. Fig.~\ref{fig:SD_lf_vs_N} shows the square root of the CR bound as well as the RMS error in $\tilde{\lambda}_f$, obtained by simulations, for different traffic parameters and $T=50$\,s, as a function of the number of samples, $N$. The estimation error reaches an asymptote as $N$ is increased for a fixed observation window length, which, again, is attributed to sample correlation. The asymptotic value of the CR bound was derived as (\ref{eq;nb_5}) and its square root is presented in Fig.~\ref{fig:asym_lf} where it is clear that the asymptote decreases with increasing the observation window length. The analytical result in (\ref{eq;nb_5}) is verified through simulations where $N$ was set to a value that is higher than $T\lambda_f/u$ by orders of magnitude.

\paragraph*{Key Message}

The PU departure rate can be estimated using ML estimation where the estimation error is lower bounded due to sample correlation. The lower bound on the error can only be decreased by increasing the total observation window length.

\subsection{RMS Error in Estimating $u$ and $\lambda_f$ under Sensing Imperfections}
\label{sec:num_MSE_u_lf_se}

\begin{figure}
\centering
\subfigure[Estimation of $u$]{\includegraphics[width=\columnwidth]{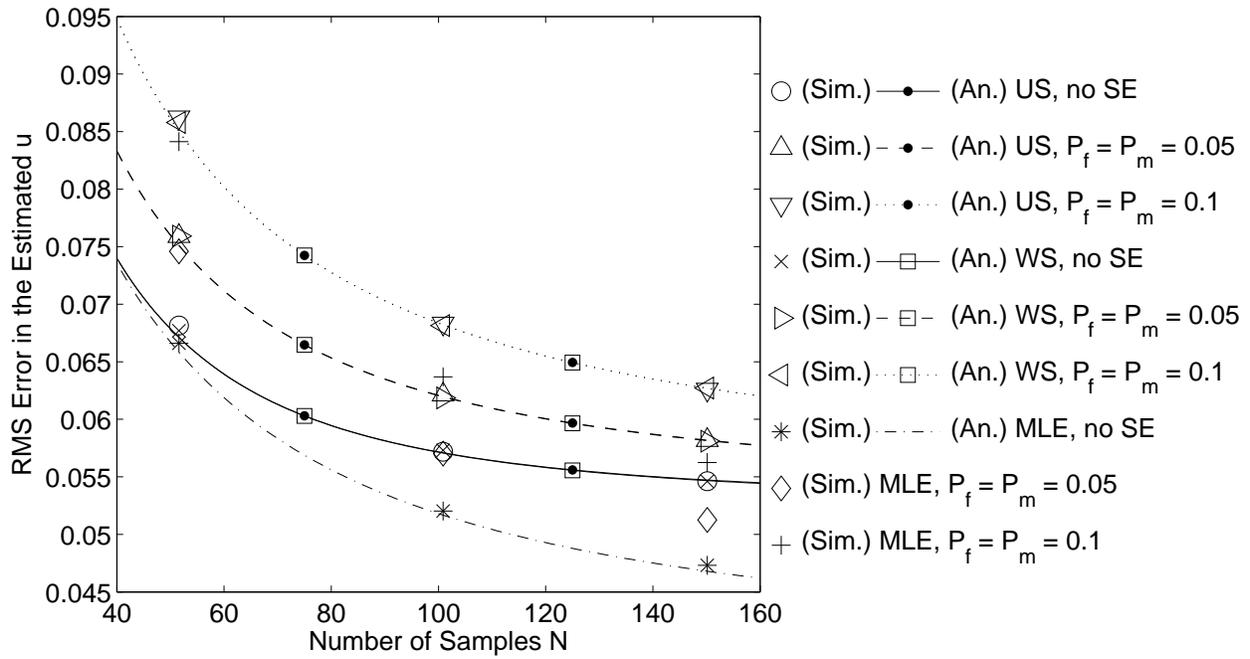}\label{fig:SD_u_vs_N_SE}}\\
\subfigure[Estimation of $\lambda_f$]{\includegraphics[width=\columnwidth]{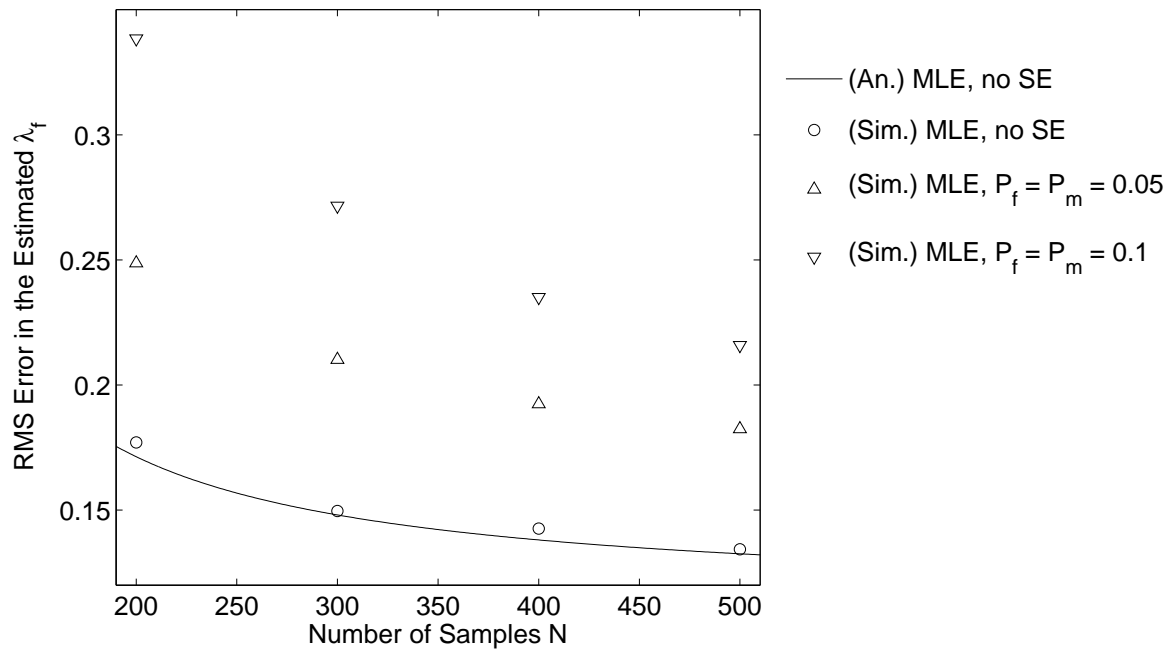}\label{fig:SD_lf_vs_N_SE}}
\caption{The impact of sensing errors (SE) on the estimation of $u$ and $\lambda_f$. The used traffic parameters: $u = 0.3$, $\lambda_f = 0.9$\,s$^{-1}$, and $T=50$\,s. The RMS error is plotted for three estimation methods; US: Averaging with uniform sampling, WS: Weighted averaging, and MLE: Maximum likelihood estimation. Simulation results (Sim.) are plotted to verify the mathematical model (An.) where applicable.}
\end{figure}

The impact of sensing imperfections on the estimation of $u$ and $\lambda_f$ is presented in Fig.~\ref{fig:SD_u_vs_N_SE} and Fig.~\ref{fig:SD_lf_vs_N_SE}, respectively. The assumed parameters are $\lambda_f = 0.9$\,s$^{-1}$, $u=0.3$, and $T=50$\,s. The estimation error is shown for $P_f=P_m=0$, $P_f=P_m=0.05$, and $P_f=P_m=0.1$. The estimation error under sensing imperfections for the averaging estimator of $u$ is expressed in closed form in (\ref{eq;nb_1_se}), and the error for the weighted averaging estimator using the optimal weighting sequence is expressed in closed form in (\ref{eq:mse_weight_opt_se}). Analysis and simulation-based results are plotted for both estimators (the special case of uniform sampling is used for the averaging estimator), where the simulation results match the theoretical expressions for both estimators. On the other hand, the impact of sensing errors on the error in the ML estimators of $u$ and $\lambda_f$ is only expressed via simulations using the modified likelihood function expressed in (\ref{eq;joint_error_ss_1}).

\paragraph*{Key Message}

For estimating $u$, the ML estimator outperforms the averaging estimators under sensing imperfections. The effect of sensing errors on the error in estimating $\lambda_f$ is noticeable where the RMS of the estimation error increases by up to 91\% for $P_f=P_m=0.1$, compared to perfect spectrum sensing. Finally, the impact of the sensing imperfections on the PU traffic parameters estimation error decreases as $N$ increases.

\subsection{Algorithm $\bf{I}$: Performance of the Proposed Duty Cycle Estimation Algorithm for Constrained $N$}
\label{sec:algorithm_numerical_cons_N}

\begin{figure}
\centering
\subfigure[]{\includegraphics[width=\columnwidth]{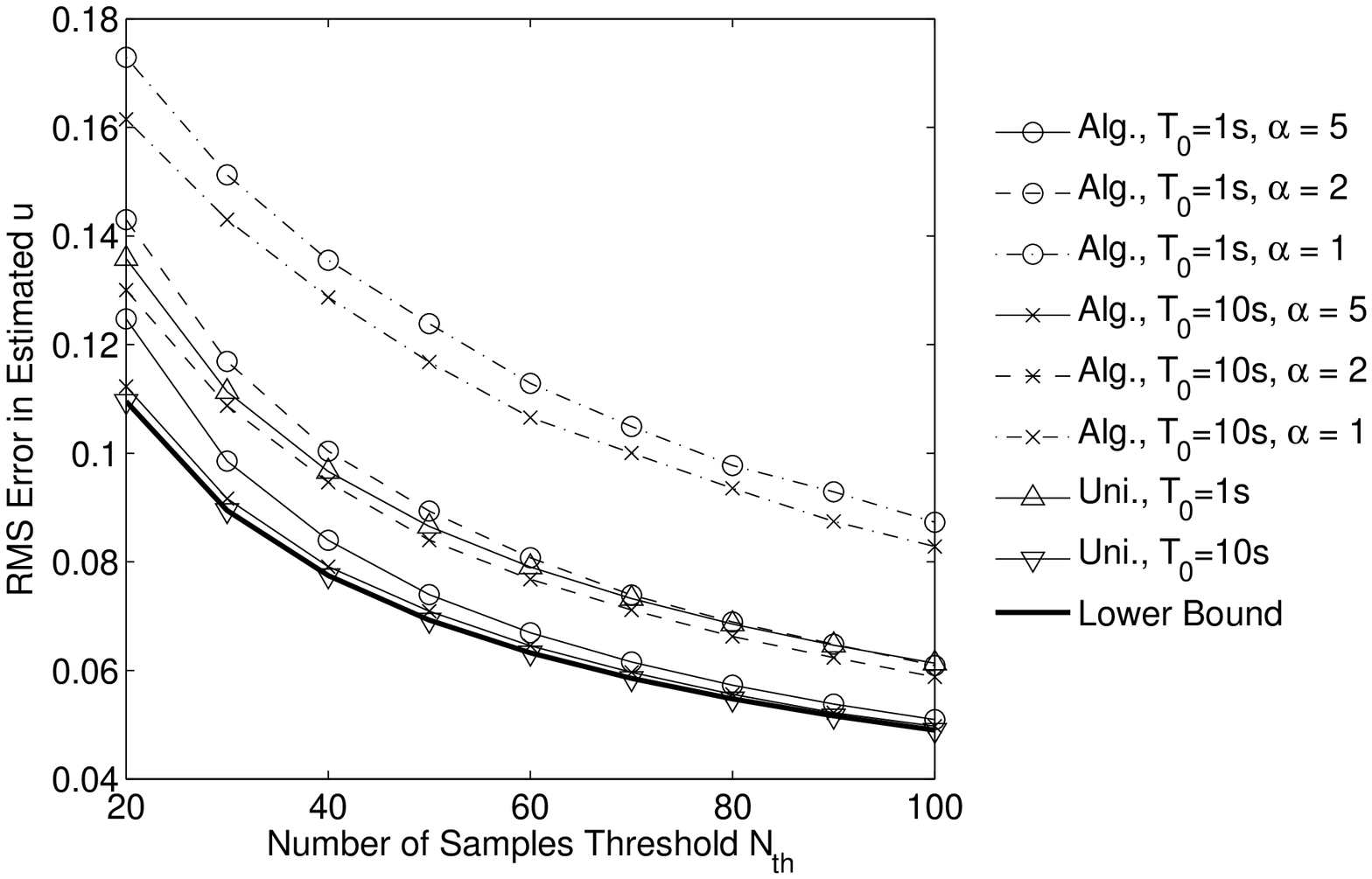}\label{fig:algo_N}}\\
\subfigure[]{\includegraphics[width=\columnwidth]{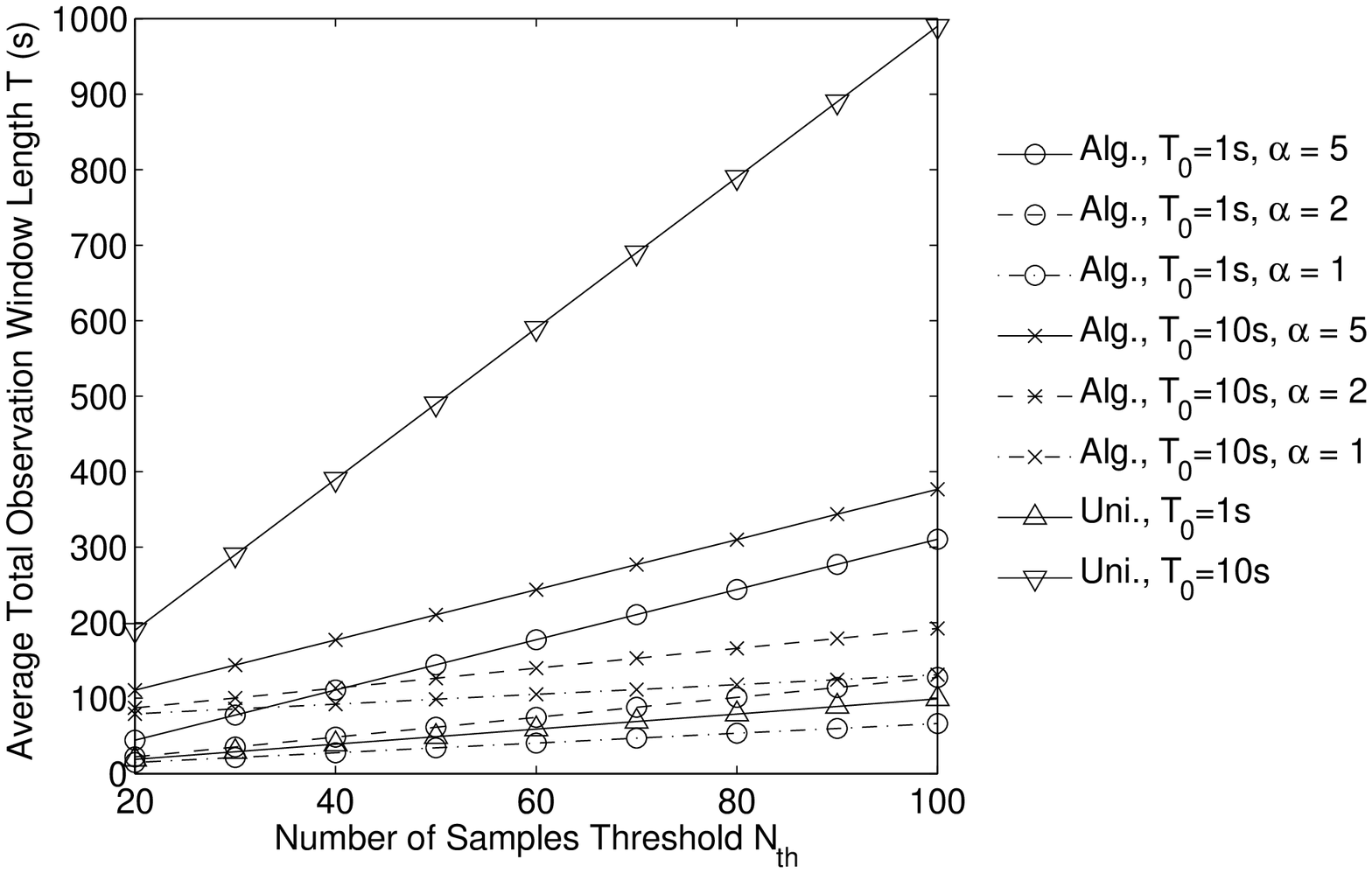}\label{fig:algo_N_time}}
\caption{The performance of Algorithm $\bf{I}$ for constrained $N_{th}$, Fig.~\ref{fig:algo_N}: The RMS estimation error in $u$ as a function of the number of samples threshold $N_{th}$; Fig.~\ref{fig:algo_N_time}: The total observation window length as a function of the number of samples threshold $N_{th}$. The estimation error and total observation window length for uniform sampling are plotted for comparison. The theoretical lower bound on error is plotted as a reference; Alg.: Algorithm $\bf{I}$, Uni.: uniform sampling. The used parameters: $u=0.6$, $\lambda_f=0.9$\,s$^{-1}$, $N_0=5$, $T_0\in\{1,10\}$\,s.}
\end{figure}

This section presents the performance of Algorithm $\bf{I}$ when the total number of samples is constrained. A typical application would be traffic estimation by an energy-constrained node. The initial number of samples, $N_0$, is set to 5 samples. The initial inter-sample time, $T_0$, is set to two different values, 1 and 10 seconds, to investigate the performance of the algorithm under different initial conditions. After the initial $N_0$ samples, the inter-sample time is adapted as described in Section~\ref{sec:algorithm}. Three different values of $\alpha\in\{1,2,5\}$, are selected to show the compromise between the estimation accuracy and the total observation window length. The performance of the algorithm is compared to that of uniform sampling with $T_u$ set equal to the 2 different values of $T_0$, i.e., the algorithm is compared to the case where the inter-sample time is kept constant at the initial conditions without adaptation. The traffic parameters are $u=0.6$ and $\lambda_f=0.9$\,s$^{-1}$. The RMS error in the estimate of $u$, $R_{\tilde{u},N}$, is presented in Fig.~\ref{fig:algo_N} and the equivalent total observation window length is presented in Fig.~\ref{fig:algo_N_time}. As a reference, the theoretical lower bound for the averaging-based estimation error as $T$ tends to infinity is also plotted. 

The results show that the algorithm can blindly achieve an RMS estimation error that is only 1.5\% higher than the theoretical lower bound for $\alpha = 5$, $T_0 = 10$\,s, and $N_{th}=100$ samples. Uniform sampling alone can achieve a low estimation error but at the expense of a notable increase in the total observation window length. Moreover, the algorithm has lower dependence on the initial conditions compared to uniform sampling, as the algorithm adapts the inter-sample time according to the traffic parameters. Furthermore, higher $\alpha$ yields a smaller estimation error but causes an increase in the estimation duration, thus, $\alpha$ can be tuned according to the application constraints.

\begin{figure*}
\centering
\subfigure[]{\includegraphics[width=0.49\columnwidth]{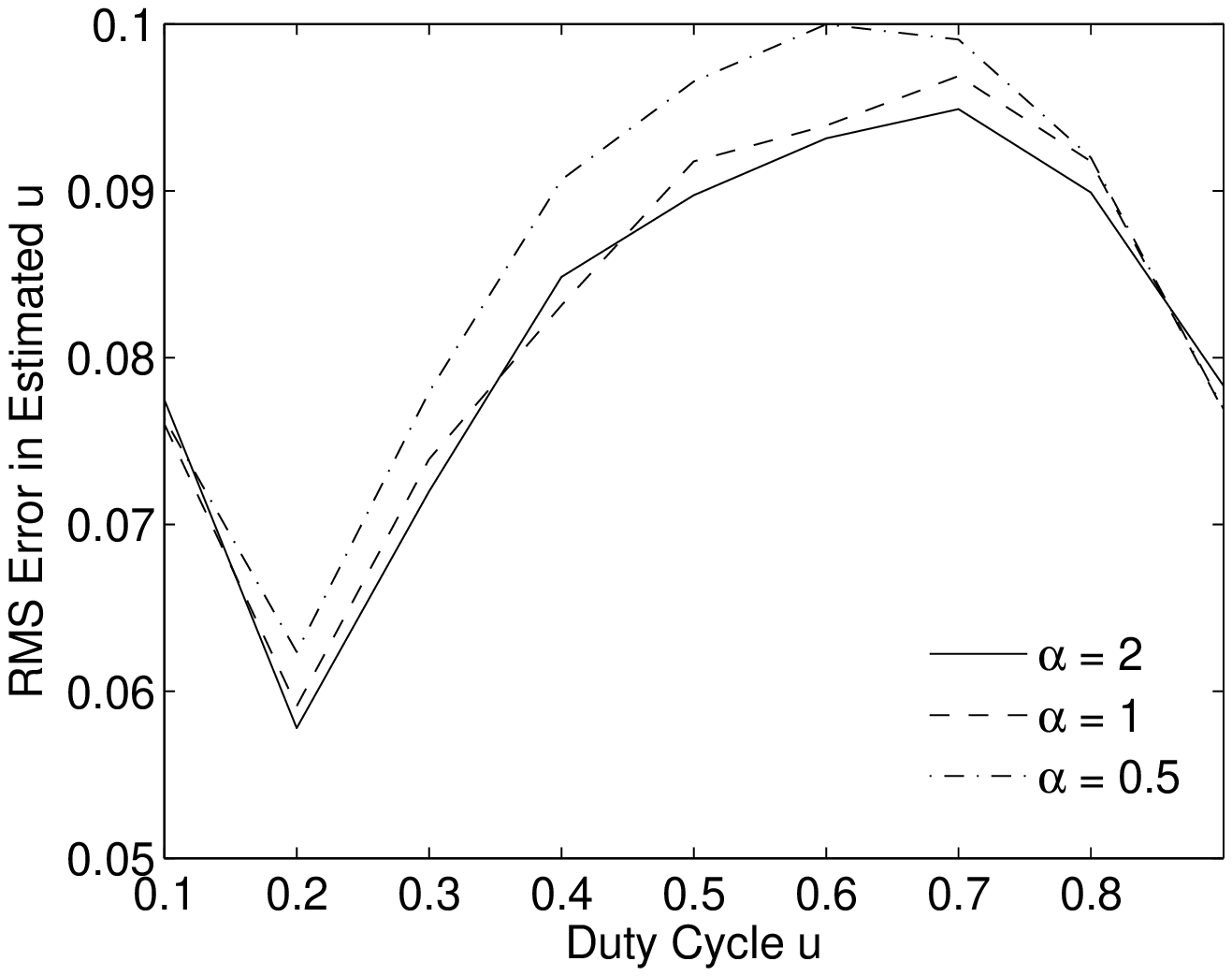}\label{fig:algo_V}}
\subfigure[]{\includegraphics[width=0.49\columnwidth]{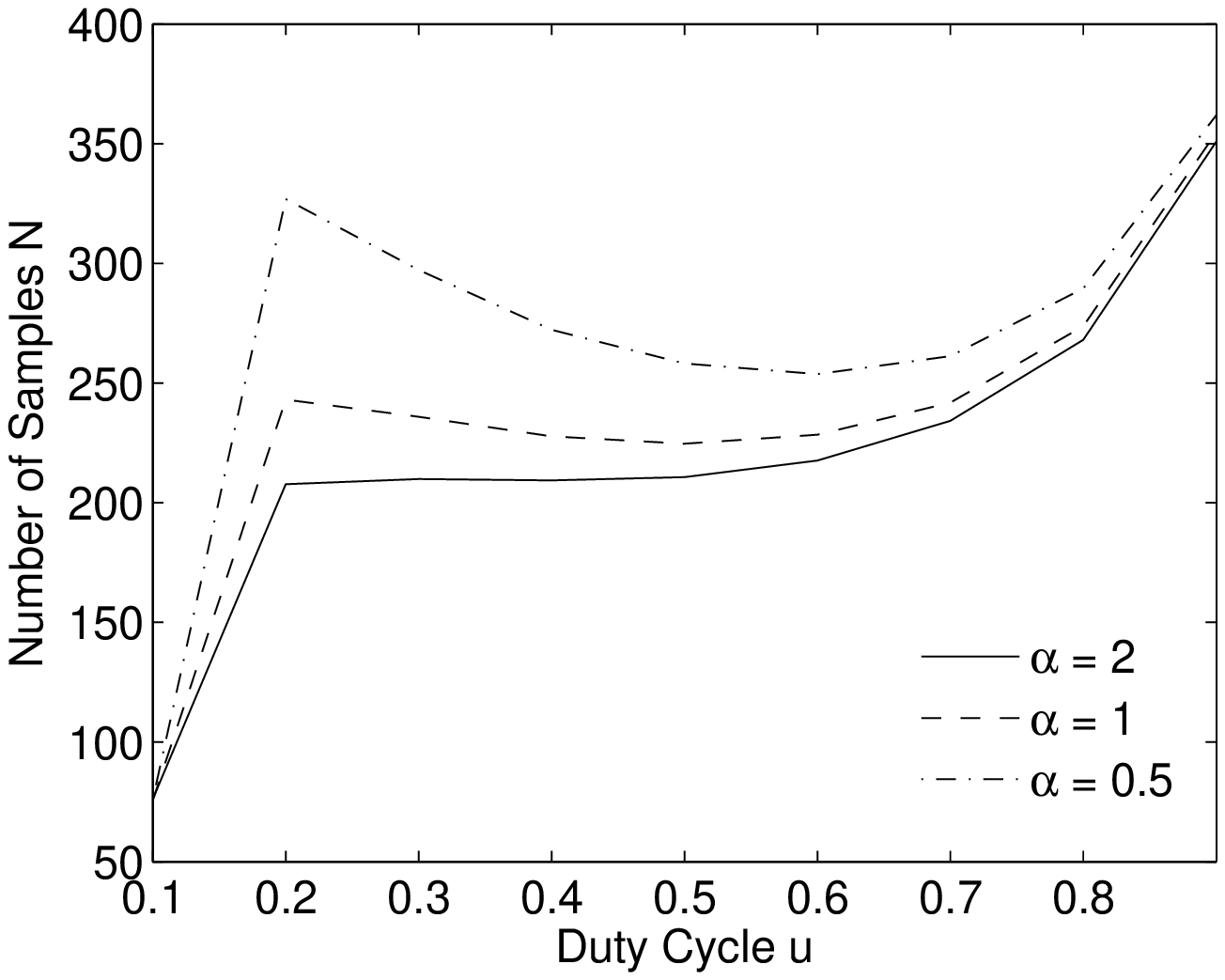}\label{fig:algo_V_N}}
\subfigure[]{\includegraphics[width=0.49\columnwidth]{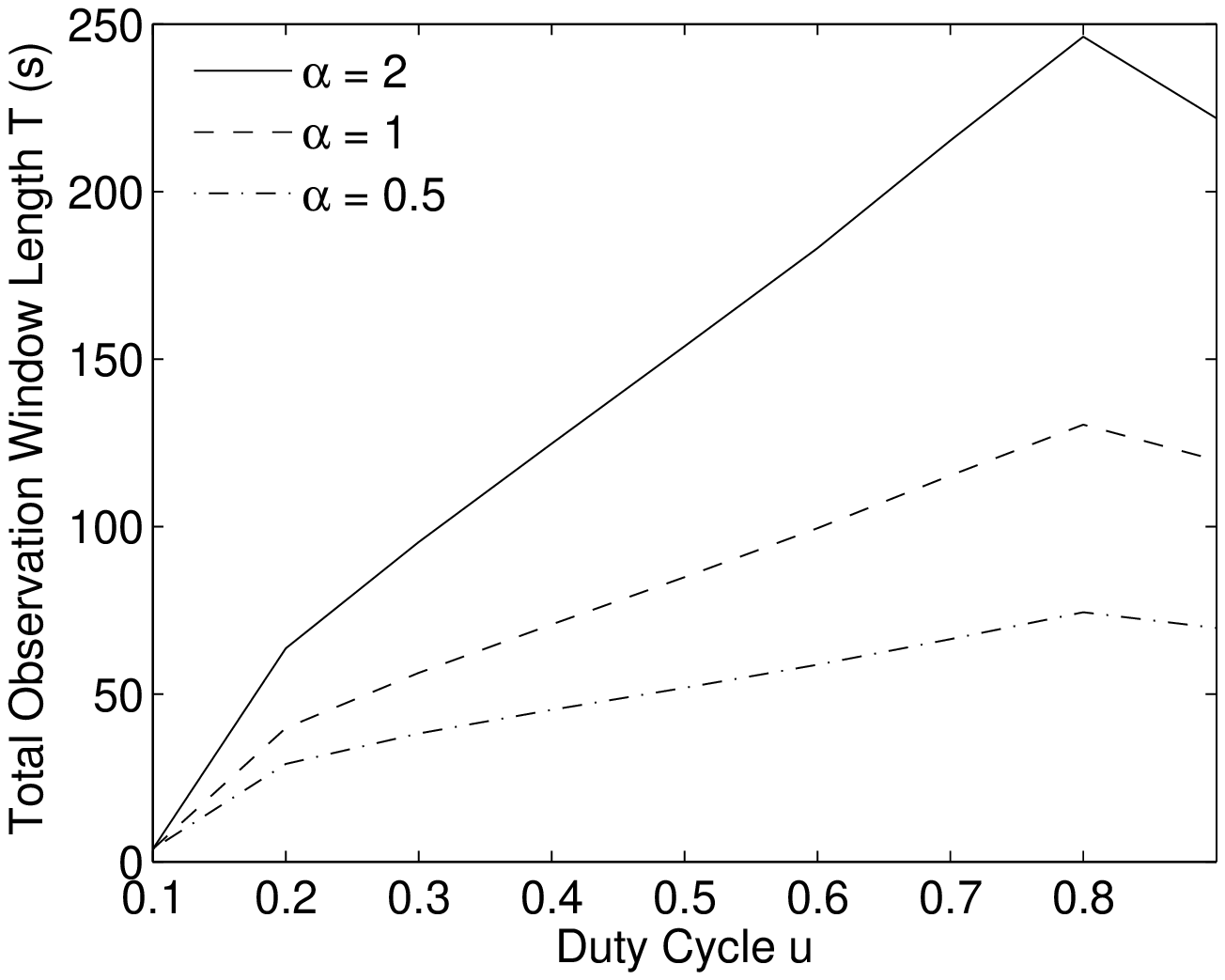}\label{fig:algo_V_T}}
\caption{The performance of Algorithm $\bf{I}$ for constrained RMS error, Fig.~\ref{fig:algo_V}: The achieved RMS estimation error in $u$ as a function of $u$; Fig.~\ref{fig:algo_V_N}: The average total number of samples as a function of $u$; Fig.~\ref{fig:algo_V_T}: The average total observation window length as a function of $u$. The used parameters: $\lambda_f=0.9$\,s$^{-1}$, $N_0=50$, $T_0=50$\,ms.}
\end{figure*}

\paragraph*{Key Message}

The proposed algorithm can blindly estimate $u$ for a constrained $N$ by adapting the inter-sample time according to the estimated $u$. The compromise between the estimation error and the total observation window length can be controlled by tuning the algorithm parameter $\alpha$.

\subsection{Algorithm $\bf{I}$: Performance of the Proposed Duty Cycle Estimation Algorithm given a Target Estimation Error}
\label{sec:algorithm_numerical_cons_V}

In this section, the duty cycle is blindly estimated until the RMS error reaches a targeted value using Algorithm $\bf{I}$. The target RMS estimation error is 0.1 and the algorithm is tested for $u$ ranging from 0.1 to 0.9, while $\lambda_f$ is set to 0.9\,s$^{-1}$. The algorithm parameters are set to $N_0=50$ samples, $T_0=50$ ms, and $\alpha\in\{1,2,5\}$. The RMS estimation error, the average total number of samples, and the average total observation window length are plotted in Fig.~\ref{fig:algo_V}, Fig.~\ref{fig:algo_V_N} and Fig.~\ref{fig:algo_V_T}, respectively.

\paragraph*{Key Message}

The achieved RMS estimation error is always reached for all values of $u$. The reached error is less than the targeted error for most cases as the algorithm targets the worst case error. Furthermore, the results emphasize the compromise between the number of samples and the observation window length where $\alpha$ can be chosen according to the sensing energy and sensing delay constraints.

\subsection{Algorithm $\bf{II}$: Joint Estimation of $u$, and $\lambda_f$ (or $\lambda_n$) for a Target Estimation Error}
\label{sec:algorithm_numerical_u_lf_ln}

\begin{figure}
\centering
\subfigure[]{\includegraphics[width=0.49\columnwidth]{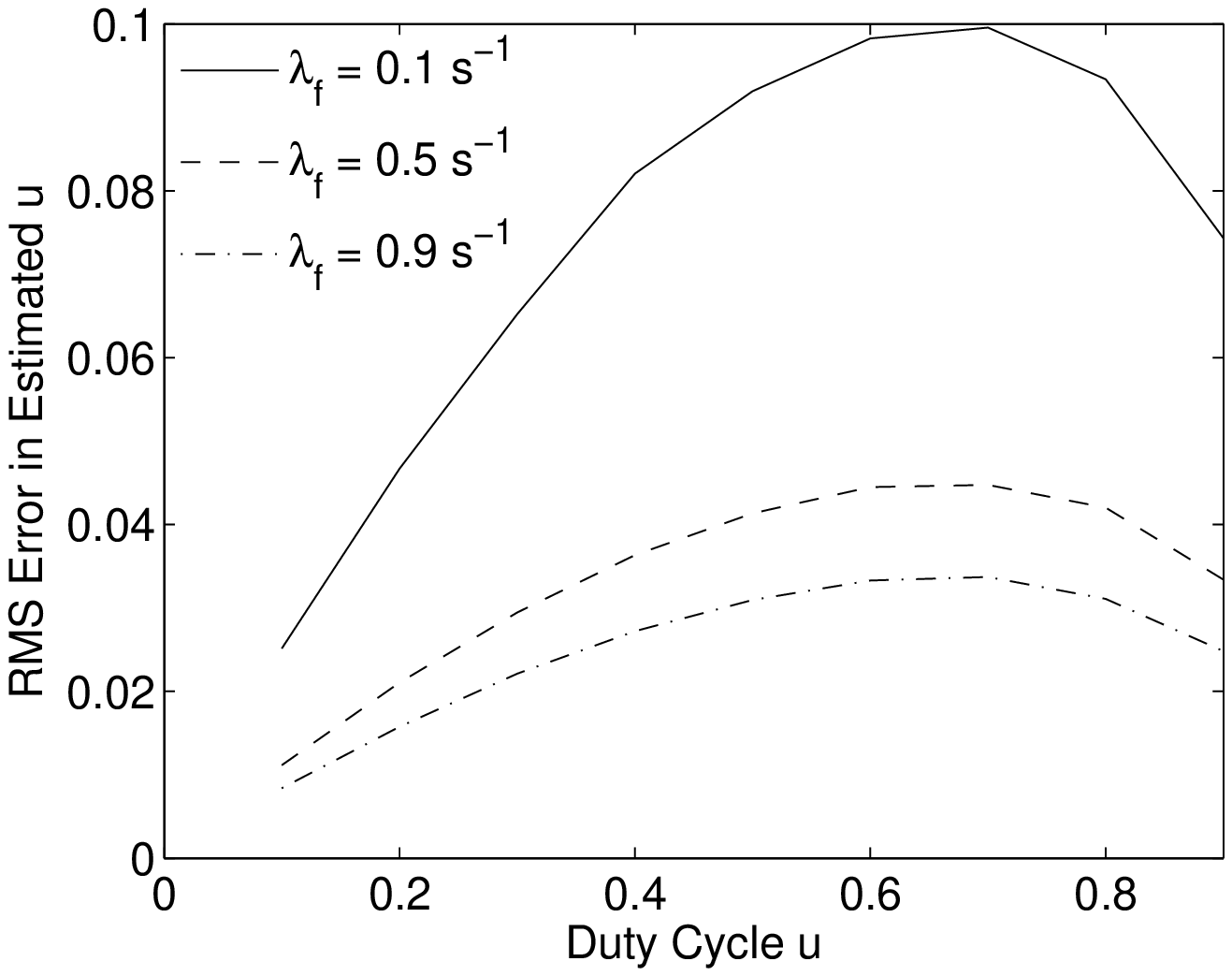}\label{fig:alg_u_lf_ln_err_u}}
\subfigure[]{\includegraphics[width=0.49\columnwidth]{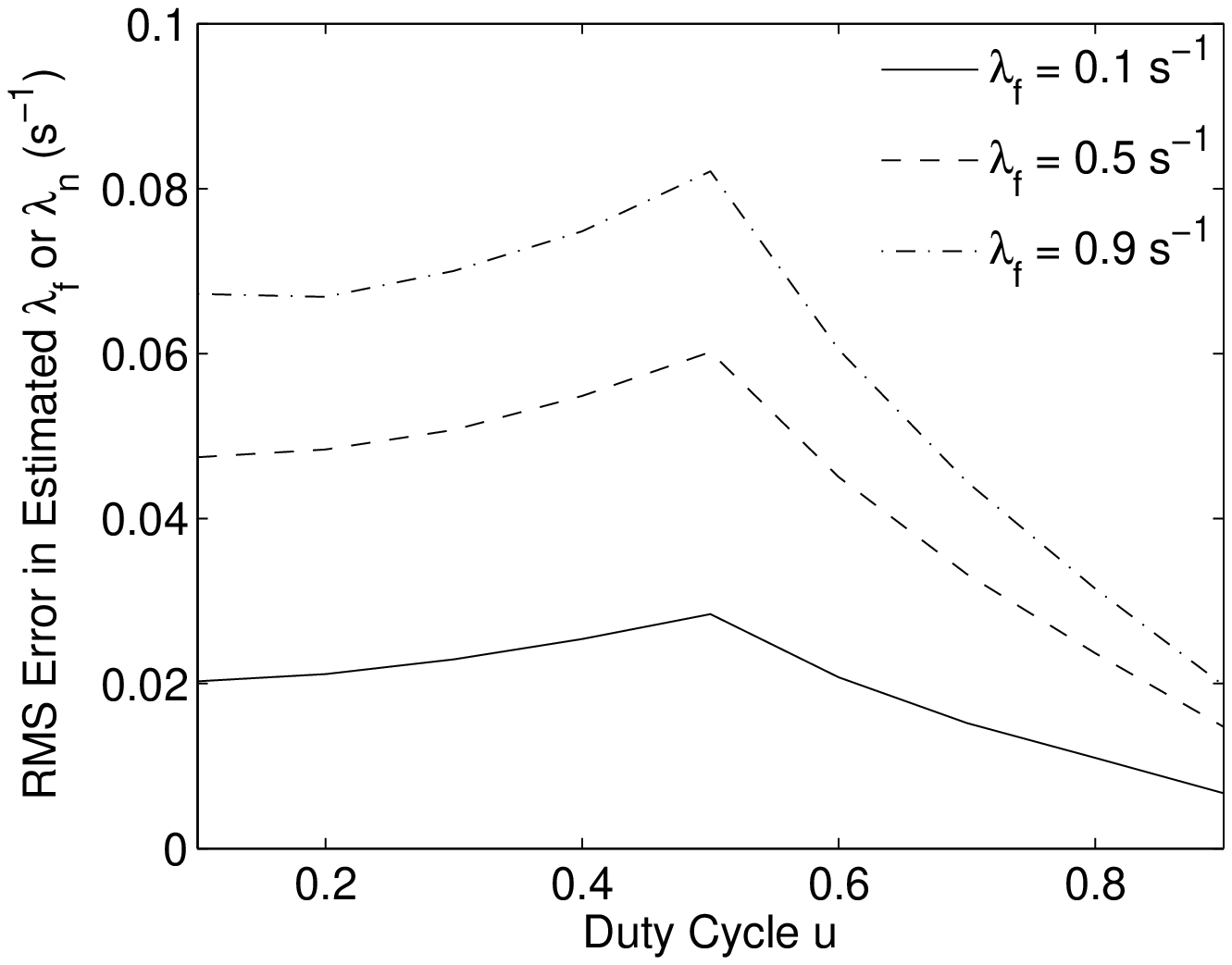}\label{fig:alg_u_lf_ln_err_lf}}
\caption{The performance of Algorithm $\bf{II}$ for the joint estimation of $u$, and $\lambda_f$ (or $\lambda_n$): Fig.~\ref{fig:alg_u_lf_ln_err_u}: The RMS estimation error in $u$ as a function of $u$; Fig.~\ref{fig:alg_u_lf_ln_err_lf}: The RMS error in $\tilde{\lambda}_f$ or $\tilde{\lambda}_n$ as a function of $u$. The used parameter: $T_0=50$\,ms.}
\end{figure}

The targeted RMS estimation error in $u$ is set to 0.1, and the targeted RMS error in $\tilde{\lambda}_f$ or $\tilde{\lambda}_n$ is set to 0.1\,s$^{-1}$. The algorithm is tested for $u$ ranging from 0.1 to 0.9, and $\lambda_f \in \{0.1,0.5,0.9\}$\, s$^{-1}$. Parameters $\lambda_{\min}$ and $\lambda_{\max}$ are set to 0.1\,s$^{-1}$ and 1\,s$^{-1}$, respectively, and the inter-sample time, $T_0$, is set to 50\,ms. The RMS estimation error in $u$ is presented in Fig.~\ref{fig:alg_u_lf_ln_err_u}, and the RMS error in $\tilde{\lambda}_f$ or $\tilde{\lambda}_n$ is shown in Fig.~\ref{fig:alg_u_lf_ln_err_lf}. The results show that the constraints on the estimation error are reached for all of the tested values of $u$ and $\lambda_f$. The estimation error in $u$ is higher for lower $\lambda_f$ due to the increased sample correlation, while the error in $\tilde{\lambda}_f$ increases with $\lambda_f$ which complies with Section~\ref{sec:lambda_f_estimation_error}. Furthermore, since the inter-sample time is constant and the algorithm terminates when the worst case error (covering the full considered range of the traffic parameters) matches the target error, the total observation window length and number of samples are constant for all considered values of $u$ and $\lambda_f$. For this specific setup, the total observation window equals 290 seconds, which is higher than that reported for Algorithm $\bf{I}$, as shown in Fig.~\ref{fig:algo_V_T}, where $\lambda_f$ was assumed to be known a priori.

\paragraph*{Key Message}

The proposed algorithm can blindly estimate all traffic parameters while satisfying a targeted estimation error.

\begin{figure*}[!t]
\normalsize
\setcounter{MYtempeqncnt}{\value{equation}}
\setcounter{equation}{40}
\begin{equation}
\bar{h}_{n,i,j} = 
\begin{cases}
\hat{h}_{n,i,j} = \sum_{a=1}^{N-1} \prod_{k=1}^{a} \Gamma_k, & i=j=1,\\
\hat{h}_{n,i,j-1} - \hat{h}_{n,i,j} = \prod_{k=1}^{j-1} \Gamma_k, & i=1,j>1,\\
\hat{h}_{n,i,j} + \hat{h}_{n,i-1,j-1} - \hat{h}_{n,i-1,j} -\hat{h}_{n,i,j-1} = -\prod_{k=i}^{j-1} \Gamma_k, & j > i>1,\\
\hat{h}_{n,i,j} + \hat{h}_{n,i-1,j-1} - \hat{h}_{n,i-1,j} -\hat{h}_{n,i,j-1} = \sum_{a=i}^{N-1} \prod_{k=i}^{a} \Gamma_k+ \sum_{a=1}^{i-1} \prod_{k=a}^{i-1} \Gamma_k, & j = i>1,\\
\bar{h}_{n,j,i}, & \text{otherwise}.\\
\end{cases}
\label{eq:h_elements}
\end{equation}
\setcounter{equation}{\value{MYtempeqncnt}}
\hrulefill
\vspace*{4pt}
\end{figure*}

\section{Conclusions}
\label{sec:conclusions}

In this paper, we developed a mathematical framework that quantifies the estimation accuracy of PU traffic parameters in the form of the mean squared estimation error. We derived the Cram\'{e}r-Rao bound on the accuracy of estimating the mean PU duty cycle, $u$, using uniform sampling under perfect knowledge of either the mean PU off- or on-time. We also analyzed the estimation error in $u$ for a variety of estimators based on sample stream averaging and maximum likelihood estimation. We proved that for the equal-weighted averaging-based estimation of $u$, the estimation error is convex with respect to the inter-sample time sequence and we derived the optimal sequence and the corresponding error. Moreover, we derived the optimal weighting sequence when weighted averaging is employed. Furthermore, we showed that the maximum likelihood estimation of $u$ outperforms all averaging based estimators, provided that a priori knowledge of the mean PU off-time or the mean PU on-time is available. Regarding the mean PU off- and on-times, we formulated the estimation error bounds, for all unbiased estimators, in the form of the Cram\'{e}r-Rao bounds. We showed that the estimation error for all PU traffic parameters for a fixed observation length is lower bounded due to sample correlation, where the bound can be reduced by increasing the total observation window length. Besides, we demonstrated the impact of spectrum sensing errors on the estimation accuracy for all PU traffic parameters, including analytical results where applicable, and showed that the effect of sensing errors on the estimation accuracy of $u$ can be eliminated by increasing the number of traffic samples. Finally, we proposed algorithms, based on the derived error expressions, for the blind estimation of $u$ for constrained number of samples, observation window length, and expected estimation error. We concluded the paper by proposing an algorithm for the joint estimation of all traffic parameters that successfully achieves a target estimation error.

\appendices

\renewcommand{\theequation}{\thesection.\arabic{equation}}

\section{Proof of $V_{\tilde{u}_{a,s},N}$}
\label{sec:math_ind_vu}

\begin{IEEEproof}The expression presented in (\ref{eq;nb_1_se}) can be proved by mathematical induction. First assume, without loss of generality, that $1-P_f-P_m\neq0$. Define $P_n = \Pr(\tilde{\boldsymbol{z}} = \tilde{\mathcal{Z}}_i|\mathcal{T})$, i.e., the probability of observing the estimated PU traffic samples vector $\tilde{\mathcal{Z}}_i$ given $\mathcal{T}$. Moreover, define $P$ as a vector containing all values of $P_n$ such that $P_n$ is the $n$th element of $P$. Furthermore, define $\mathcal{S}$ as a vector where $S_{i}$, defined in Section~\ref{sec:estimation_u_err}, is the $i$th element of $\mathcal{S}$. Hence, $V_{\tilde{u}_{a,s},N} = \sum_{i=1}^{2^N} S_i^2 P_i - u^2$.

For the base case with $N=2$, $\tilde{\mathcal{Z}} = [00,01,10,11]$ and $\mathcal{S} = -1/(1-P_f-P_m)[-P_f,-P_f+1/2,-P_f+1/2,-P_f+1]$. Besides, $P = [(1-u)\bar{P}_f(\textstyle\Pr_{00}\bar{P}_f+\textstyle\Pr_{01}P_m) + uP_m(\textstyle\Pr_{10}\bar{P}_f+\textstyle\Pr_{11}P_m), (1-u)\bar{P}_f(\textstyle\Pr_{00}P_f+\textstyle\Pr_{01}\bar{P}_m) + uP_m(\textstyle\Pr_{10}P_f+\textstyle\Pr_{11}\bar{P}_m), (1-u)P_f(\textstyle\Pr_{00}\bar{P}_f+\textstyle\Pr_{01}P_m) + u\bar{P}_m(\textstyle\Pr_{10}\bar{P}_f+\textstyle\Pr_{11}P_m), (1-u)P_f(\textstyle\Pr_{00}P_f+\textstyle\Pr_{01}\bar{P}_m) + u\bar{P}_m(\textstyle\Pr_{10}P_f+\textstyle\Pr_{11}\bar{P}_m)]$, where $\bar{P}_f = 1-P_f$ and $\bar{P}_m = 1-P_m$. Accordingly, \addtocounter{equation}{1}
\begin{align}
V_{\tilde{u}_{a,s},2}&=\frac{u(1-u)\Gamma_c}{2} + \frac{u(1-u)}{2}\nonumber\\&\quad + \frac{uP_m(1-P_m)+(1-u)P_f(1-P_f)}{2(1-P_f-P_m)^2},
\end{align}
which equals (\ref{eq;nb_1_se}) for $N=2$, hence, proves the base case.

Showing that (\ref{eq;nb_1_se}) holds for $N+1$ while assuming that it is true for $N$ is sufficient for proving (\ref{eq;nb_1_se}). Subscripts $N$ and $N+1$ are added to $P_n$ and $S_n$ to differentiate between cases with $N$ and $N+1$ samples. For $N+1$ samples, $V_{\tilde{u}_{a,s},N+1} = \sum_{i=1}^{2^{N+1}} S_{i,N+1}^2 P_{i,N+1} - u^2$ and can be expressed as
\begin{equation}
V_{\tilde{u}_{a,s},N+1} = \frac{N^2 V_{\tilde{u}_{a,s},N}}{(N+1)^2} + \Omega_1 + \Omega_2 - \frac{(2N+1)u^2}{(N+1)^2},
\label{eq:proof_4_1}
\end{equation}
where
\begin{align}
\Omega_1 &= \sum_{i=1}^{2^{N+1}} P_{i,N+1} \left(\sum_{n=1}^{N}\tilde{\mathcal{Z}}_{i,n}-N P_f\right)\left(\tilde{\mathcal{Z}}_{i,N+1}-P_f\right)\nonumber\\&\quad\times \frac{2}{(N+1)^2(1-P_f-P_m)^2}
\end{align}
and
\begin{align}
\Omega_2 & = \sum_{i=1}^{2^{N+1}} P_{i,N+1} (\tilde{\mathcal{Z}}_{i,N+1}-P_f)^2\nonumber\\&\quad\times \frac{1}{(N+1)^2(1-P_f-P_m)^2}.
\end{align}
Note that $\tilde{\mathcal{Z}}_{i,N+1}$ in $\Omega_1$ and $\Omega_2$ denotes the estimated traffic sample number $N+1$ in the traffic sample vector $\tilde{\mathcal{Z}_i}$. $\Omega_1$ is a recursive expression that can be simplified to
\begin{equation}
\Omega_1 = \frac{2Nu^2}{(N+1)^2} + \frac{2u(1-u)}{(N+1)^2}\sum_{j=1}^{N} \prod_{k=j}^{N} e^{\frac{-T_{k}\lambda_f}{u}},
\end{equation}
where the proof is omitted for brevity. Moreover, $\Omega_2$ can be simplified to
\begin{equation}
\Omega_2 = \frac{P_f(1-P_f)+u(1-2P_f)(1-P_f-P_m)}{(N+1)^2(1-P_f-P_m)^2}.
\end{equation}
Finally, substituting (\ref{eq;nb_1_se}) for $V_{\tilde{u}_{a,s},N}$ in (\ref{eq:proof_4_1}) and using the simplified expressions for $\Omega_1$ and $\Omega_2$, we obtain
\begin{align}
V_{\tilde{u}_{a,s},N+1} & = \frac{2u(1-u)}{(N+1)^2} \sum_{i=1}^{N} \sum_{j=1}^{N+1-i} \prod_{k=j}^{i+j-1} e^{\frac{-T_{k}\lambda_f}{u}} + \frac{u(1-u)}{N+1}\nonumber\\ & + \frac{uP_m\left(1-P_m\right) + \left(1-u\right)P_f\left(1-P_f\right)}{(N+1)\left(1-P_f-P_m\right)^2}.
\end{align}
This corresponds to (\ref{eq;nb_1_se}) with the number of samples set to $N+1$, hence, proves (\ref{eq;nb_1_se}) by mathematical induction.\end{IEEEproof}

\section{Proof of Convexity of $V_{\tilde{u}_a,N}\left( \mathcal{T} \right)$}
\label{sec:V_u_conv}

\begin{IEEEproof}
In this section, expression (\ref{eq;nb_1}) is proved to be convex by showing that the Hessian of $V_{\tilde{u}_a,N}\left( \mathcal{T} \right)$, $\nabla^2 V_{\tilde{u}_a,N}\left( \mathcal{T} \right)$, is positive-semidefinite~\cite{berghe}. Denote the Hessian of $V_{\tilde{u}_a,N}\left( \mathcal{T} \right)$ by $H = [h_{a,b}]$. $H$ is a symmetric $\{N-1\}$-by-$\{N-1\}$ matrix and can be expressed as $h_{a,b} = \frac{2u(1-u)}{N^2} \left(\frac{\lambda_f}{u}\right)^2 \sum_{i=1}^{a} \sum_{j=b}^{N-1} \prod_{k=i}^{j} \Gamma_k$, $\forall a \leq b$. Since $\frac{2u(1-u)}{N^2} \left(\frac{\lambda_f}{u}\right)^2 \geq 0$, showing that $\hat{H}$ is positive-semidefinite proves that $H$ is positive-semidefinite, where $\hat{H} = \frac{N^2}{2u(1-u)} \left(\frac{u}{\lambda_f}\right)^2 H$. $\hat{H}$ is proved to be positive-semidefinite by showing that all of its $N-1$ leading principal minors are non-negative. Define $\hat{H}_n = [\hat{h}_{n,i,j}]$ as the upper left $n$-by-$n$ corner of $\hat{H}$. Showing that the determinant of $\hat{H}_n$ is non-negative for $n \in \{1,2,\cdots N-1\}$ proves that $\hat{H}$ is positive-semidefinite. Define $\bar{H}_n = [\bar{h}_{n,i,j}]$ as a symmetric matrix where the elements of $\bar{H}_n$ are defined in (\ref{eq:h_elements}), given at the top of the previous page. 

It is easy to show that $|\bar{H}_n| = |\hat{H}_n|$ as $\bar{H}_n$ is formed by performing row and column addition operations on $\hat{H}_n$ and an even number of sign changes. Moreover, $\bar{H}_n$ is diagonally dominant, that is, for every row of the matrix, the magnitude of the diagonal element is greater than or equal to the summation of the magnitude of the other non-diagonal elements: $|\bar{h}_{n,i,i}| - \sum_{j \neq i} |\bar{h}_{n,i,j}| = \sum_{a=n}^{N-1} \prod_{k=i}^{a} \Gamma_k \geq 0 , \forall i \in \{1,\cdots,n\}, \forall n \in \{1,\cdots,N-1\}$. A real symmetric diagonally dominant matrix with non-negative diagonal elements is positive-semidefinite~\cite[Ch. 6]{matrix_book}. Hence, $|\bar{H}_n| \geq 0$, and accordingly, $|\hat{H}_n| \geq 0$, for $n \in \{1,2,\cdots N-1\}$. Consequently, $\hat{H}$ is positive-semidefinite and $V_{\tilde{u}_a,N}\left( \mathcal{T} \right)$ is convex.
\end{IEEEproof}

\section{Proof of the Lower Bound on $V_{\tilde{u}_m,N}$}
\label{sec:math_ind_u}

\begin{IEEEproof} We first start by simplifying the expression for the Fisher information. Let $\Phi_N = \partial \log L(\boldsymbol{z}|u)/\partial u$ where $N$ samples are used for estimation. Note that $I_m\left(u,N\right) = E\left[\Phi_N^2\right]$. Using (\ref{eq:p_xy}) and (\ref{eq:likelihood_func_lambda_f2}), $\Phi_N$ can be written as
\begin{align}
\Phi_N & = \frac{z_1-u}{u(1-u)} + \Phi^0 \left[\frac{n_1}{\Pr_{01}(T_c|u)}-\frac{n_0}{\Pr_{00}(T_c|u)}\right]\nonumber\\&\quad + \Phi^1 \left[\frac{n_3}{\Pr_{11}(T_c|u)}-\frac{n_2}{\Pr_{10}(T_c|u)}\right],
\label{eq:derv_u_ML_1}
\end{align}
where $\Phi^0 = \left(1-\Gamma_c\right)-\lambda_f T_c \Gamma_c/u$, $\Phi^1 = \left(1-\Gamma_c\right)+\lambda_f T_c \Gamma_c \left(1-u\right)/u^2$, and $\Gamma_c = e^{-\lambda_f T_c/u}$. We now apply mathematical induction to prove (\ref{eq;nb_4_u}). Starting with the base case of $N = 2$, $I_m\left(u,2\right) = E\left[\Phi_2^2\right]$. The expectation is evaluated over all four possible sample sequences. Thus, $I_m\left(u,2\right) = \sum_{i=1}^{4} \phi_{2,i}^2 \Pr(\boldsymbol{z} = \mathcal{Z}_i|\mathcal{T})$, where $\phi_{N,i}$ is $\Phi_N$ evaluated for $\mathcal{Z}_i$. Furthermore, $\mathcal{Z} = [00,01,10,11]$ and the corresponding values of $\Pr(\boldsymbol{z} = \mathcal{Z}_i|\mathcal{T}) = [(1-u)\Pr_{00}(T_1), (1-u)\Pr_{01}(T_1), u\Pr_{10}(T_1), u\Pr_{11}(T_1)]$, where $\Pr_{xy}(\cdot)$ is as defined in (\ref{eq:p_xy}). Denote the vector of all values of $\phi_{2,i}$ for $i \in \{1,2,3,4\}$ by $\phi_{2}$. Hence, $\phi_{2} = \left[ \frac{1}{(u-1)} - \frac{u\left(1-\Gamma_c\right)-\lambda_f T_c \Gamma_c}{u\Pr_{00}(T_c|u)}, \frac{1}{(u-1)} + \frac{u\left(1-\Gamma_c\right)-\lambda_f T_c \Gamma_c}{u\Pr_{01}(T_c|u)},\right.$ $\left.\frac{1}{u} - \frac{u^2\left(1-\Gamma_c\right)+\lambda_f T_c \Gamma_c \left(1-u\right)}{u^2\Pr_{10}(T_c|u)},\frac{1}{u} + \frac{u^2\left(1-\Gamma_c\right)+\lambda_f T_c \Gamma_c \left(1-u\right)}{u^2\Pr_{11}(T_c|u)}\right]$. \\Accordingly, $I_m\left(u,2\right)$ can be written as
\begin{align}
I_m\left(u,2\right) & = \frac{M_{2,1}+M_{2,2}+M_{2,3}}{(1 - \Gamma_c)(\Gamma_c + u - u\Gamma_c)(u\Gamma_c - u + 1)}\nonumber\\&\quad\times\frac{1}{u^3(1-u)},
\label{sec:derv_u_ML_2}
\end{align}
where $M_{2,1}=\lambda_f T_c\Gamma_c^2(1 - u)\left[\lambda_f T_c(1 - u)(1+\Gamma_c)\right.-$ $\left.2u(1 - 2u)(1-\Gamma_c)\right]$, $M_{2,2}=\Gamma_c^3 u^2\left[4u(u - 1) + 1\right] - \Gamma_c^2 u^2\left[10u(u - 1) + 3\right]$, and $M_{2,3}=\Gamma_c u^2\left[2(5u^2 - 5u \right.+$ $\left. 1) + 2u(1 - u)\right] + 2u^3(1 - u)$. Expression (\ref{sec:derv_u_ML_2}) is equivalent to (\ref{eq;nb_4_u}) for $N=2$, which proves the base case. Assuming that (\ref{eq;nb_4_u}) is true for any $N$, proving that (\ref{eq;nb_4_u}) holds for $N+1$ completes the proof. The Fisher information for $N+1$, can be expressed as
\begin{align}
I_m\left(u,N+1\right) & = \sum_{i=1}^{2^{N+1}} \phi_{N+1,i}^2\nonumber\\&\quad\times \Pr(\boldsymbol{z}^{(N+1)} = \mathcal{Z}_i^{(N+1)}|\mathcal{T}^{(N+1)}),
\label{eq:derv_u_ML_3}
\end{align}
where the superscripts $(N)$ and $(N+1)$ are appended to $\boldsymbol{z}$, $\mathcal{Z}_n$, $\mathcal{Z}$, and $\mathcal{T}$, to differentiate between the cases with $N$ and $N+1$ samples, respectively. The set of all possible sample sequences of length $N$, $\mathcal{Z}^{(N)}$, can be split to two subsets, $\mathcal{Z}^{(N,0)}$ and $\mathcal{Z}^{(N,1)}$, which represent the set of all sample sequences ending with $0$ and $1$, respectively. Thus, the summation over the set $\mathcal{Z}^{(N+1)}$ presented in (\ref{eq:derv_u_ML_3}) can be split to summations over the sets $\mathcal{Z}^{(N,0)}$ and $\mathcal{Z}^{(N,1)}$ as follows
\begin{align}
I_m&(u, N\!\!+\!1)\! \nonumber\\&= \sum_{\mathcal{Z}_{i}^{(N)} \in \mathcal{Z}^{(N,0)}} \textstyle\Pr_{i,N}\bigg[\left(\phi_{N,i}-\frac{\Phi^0}{\textstyle\Pr_{00}(T_c|u)}\right)^2 \textstyle\Pr_{00}(T_c|u)\nonumber\\& + \left(\phi_{N,i}+\frac{\Phi^0}{\textstyle\Pr_{01}(T_c|u)}\right)^2 \textstyle\Pr_{01}(T_c|u)\bigg] \nonumber \\
& + \sum_{\mathcal{Z}_{i}^{(N)} \in \mathcal{Z}^{(N,1)}} \textstyle \Pr_{i,N} \bigg[\left(\phi_{N,i}-\frac{\Phi^1}{\textstyle\Pr_{10}(T_c|u)}\right)^2 \textstyle\Pr_{10}(T_c|u)\nonumber\\& + \left(\phi_{N,i}+\frac{\Phi^1}{\textstyle\Pr_{11}(T_c|u)}\right)^2 \textstyle\Pr_{11}(T_c|u)\bigg] \nonumber \\
& = \sum_{\mathcal{Z}_{i}^{(N)} \in \mathcal{Z}^{(N,0)}} \frac{\textstyle\Pr_{i,N}{\Phi^0}^2}{u\left(1-\Gamma_c\right)\left(u\Gamma_c-u+1\right)}\nonumber\\& + \sum_{\mathcal{Z}_{i}^{(N)} \in \mathcal{Z}^{(N,1)}} \frac{\textstyle\Pr_{i,N}{\Phi^1}^2}{(1 - \Gamma_c)(1 - u)\left[u + (1-u)\Gamma_c\right]}+I_m(u, N) \nonumber \\
& = \frac{(u\Gamma_c - u + 1)\left[(1-\Gamma_c)u^2 + \lambda_f T_c\Gamma_c(1-u)\right]^2}{u^3(1-u)(1-\Gamma_c)(\Gamma_c + u - u\Gamma_c)(u\Gamma_c - u + 1)}\notag\\
&\quad +\frac{(1 - u)^2(\Gamma_c + u - u\Gamma_c)\left[u(1-\Gamma_c) - \lambda_f T_c\Gamma_c\right]^2}{u^3(1-u)(1-\Gamma_c)(\Gamma_c + u - u\Gamma_c)(u\Gamma_c - u + 1)}\nonumber\\&\quad+ I_m(u, N),
\label{eq:derv_u_ML_4}
\end{align}
where $\Pr_{x,y}=\Pr(\boldsymbol{z}^{y} = \mathcal{Z}_{x}^{y}|\mathcal{T}^y)$. The expression for $I_m(u, N+1)$ given in (\ref{eq:derv_u_ML_4}) can be shown to match that in (\ref{eq;nb_4_u}) for $N+1$ by substituting $I_m(u, N)$ in (\ref{eq:derv_u_ML_4}) by (\ref{eq;nb_4_u}) and simplifying the resulting expression. This concludes the proof.
\end{IEEEproof}

\section{Proof of the Lower Bound on $V_{\tilde{\lambda}_f,N}$}
\label{sec:math_ind_lf}

\begin{IEEEproof} As in Appendix~\ref{sec:math_ind_u}, let $\Theta_N = \partial \log L(\boldsymbol{z}|\lambda_{f})/\partial \lambda_{f}$, where $I_m\left(\lambda_f,N\right) = E\left[\Theta_N^2\right]$. Using (\ref{eq:p_xy}), and (\ref{eq:likelihood_func_lambda_f2}) (with the condition on $u$ replaced by a condition on $\lambda_f$), $\Theta_N$ can be written as
\begin{align}
\Theta_N & = \frac{\Gamma_c T_c}{u} \left[ u \left( \frac{n_1}{\textstyle\Pr_{01}(T_c|\lambda_f)} - \frac{n_0}{\textstyle\Pr_{00}(T_c|\lambda_f)} \right)\nonumber\right.\\&\left.\quad + (1-u) \left( \frac{n_2}{\textstyle\Pr_{10}(T_c|\lambda_f)} - \frac{n_3}{\textstyle\Pr_{11}(T_c|\lambda_f)}\right) \right].
\label{eq:derv_lf_3}
\end{align}
We now apply mathematical induction to prove (\ref{eq;nb_4}). Starting with the base case of $N = 2$, $I_m\left(\lambda_f,2\right) = E\left[\Theta_2^2\right] = \sum_{i=1}^{4} \theta_{2,i}^2 \Pr(\boldsymbol{z} = \mathcal{Z}_i|\mathcal{T})$, where $\theta_{N,i}$ is $\Theta_N$ evaluated for $\mathcal{Z}_i$, and $\mathcal{Z}$ and the corresponding values of $\Pr(\boldsymbol{z} = \mathcal{Z}_i|\mathcal{T})$ are as defined in Section~\ref{sec:math_ind_u}. Denote the vector of all values of $\theta_{2,i}$ by $\theta_{2}$, hence, $\theta_{2} = \left[ \frac{- \Gamma_c T_c}{\textstyle\Pr_{00}(T_c|\lambda_f)},\frac{\Gamma_c T_c}{\textstyle\Pr_{01}(T_c|\lambda_f)}, \frac{\Gamma_c T_c (1-u)}{u\textstyle\Pr_{10}(T_c|\lambda_f)},\frac{-\Gamma_c T_c (1-u)}{u\textstyle\Pr_{11}(T_c|\lambda_f)} \right]$. Accordingly, 
\begin{equation}
I_m\left(\lambda_f,2\right)=\frac{(\Gamma_c T_c)^2(1-u)(1+\Gamma_c)}{u(1-\Gamma_c)[\Gamma_c + u(1-\Gamma_c)^2(1-u)]},
\label{sec:I_m_2}
\end{equation}
which is equivalent to (\ref{eq;nb_4}) for $N=2$. Assuming that (\ref{eq;nb_4}) is true for any $N$, proving that (\ref{eq;nb_4}) holds for $N+1$ is sufficient for proving that (\ref{eq;nb_4}) is valid for any $N$. For $N+1$, the Fisher information can be expressed as
\begin{align}
I_m(\lambda_f,N+1)& = \sum_{i=1}^{2^{N+1}} \theta_{N+1,i}^2\nonumber\\&\times \Pr(\boldsymbol{z}^{(N+1)} = \mathcal{Z}_i^{(N+1)}|\mathcal{T}^{(N+1)}),
\label{eq:derv_lf_5}
\end{align}
where the superscripts $(N)$ and $(N+1)$ are appended to $\boldsymbol{z}$, $\mathcal{Z}_n$, $\mathcal{Z}$, and $\mathcal{T}$, to differentiate between the cases with $N$ and $N+1$ samples, respectively. As in Appendix~\ref{sec:math_ind_u}, the summation in (\ref{eq:derv_lf_5}) can be split to summations over sample sequences of length $N$ ending with 0 and 1. Thus
\begin{align}
I_m&(\lambda_f, N+1)\nonumber\\& = \sum_{\mathcal{Z}_{i}^{(N)} \in \mathcal{Z}^{(N,0)}} \textstyle\Pr_{i,N}\bigg[\left(\theta_{N,i}-\frac{\Gamma_c T_c}{1-u(1-\Gamma_c)}\right)^2 \textstyle\Pr_{00}(T_c|u)\nonumber\\&\quad + \left(\theta_{N,i}+\frac{\Gamma_c T_c}{u(1-\Gamma_c)}\right)^2 \textstyle\Pr_{01}(T_c|u)\bigg] \nonumber \\
&\quad + \sum_{\mathcal{Z}_{i}^{(N)} \in \mathcal{Z}^{(N,1)}} \textstyle \Pr_{i,N} \bigg[\left(\theta_{N,i}+\frac{\Gamma_c T_c}{u(1-\Gamma_c)}\right)^2 \textstyle\Pr_{10}(T_c|u)\nonumber\\&\quad + \left(\theta_{N,i}+\frac{u-1}{u^2 + u(1-u)\Gamma_c}\right)^2 \textstyle\Pr_{11}(T_c|u)\bigg].
\label{eq:derv_lf_6}
\end{align}
Substituting (\ref{eq:p_xy}) in (\ref{eq:derv_lf_6}) and simplifying the resulting expression
\begin{align}
I_m(\lambda_f,N+1) & = \sum_{\mathcal{Z}_{i}^{(N)} \in \mathcal{Z}^{(N,0)}} \textstyle \Pr_{i,N} \frac{(\Gamma_c T_c)^2}{u(1- \Gamma_c)[1- u(1 - \Gamma_c)]}\nonumber\\&\quad + \sum_{\mathcal{Z}_{i}^{(N)} \in \mathcal{Z}^{(N,1)}} \textstyle \Pr_{i,N} \frac{(\Gamma_c T_c)^2(1-u)}{u^2\left(1-\Gamma_c\right)(\Gamma_c+u-\Gamma_c u)} \nonumber \\&+\quad I_m(\lambda_f, N)\nonumber\\& = \frac{(\Gamma_c T_c)^2(1-u)(1+\Gamma_c)}{u(1-\Gamma_c)[\Gamma_c+u(1-\Gamma_c)^2(1-u)]}\nonumber\\&\qquad+ I_m(\lambda_f, N).
\label{eq:derv_lf_7}
\end{align}
Finally, substituting (\ref{eq;nb_4}) in (\ref{eq:derv_lf_7}) yields
\begin{align}
I_m(\lambda_f,N+1) & = \frac{N(\Gamma_c T_c)^2(1+\Gamma_c)}{(1-\Gamma_c)[\Gamma_c+u(1-\Gamma_c)^2(1-u)]}\nonumber\\&\quad\times\frac{1-u}{u}.
\label{eq:derv_lf_8}
\end{align}
This proves that (\ref{eq;nb_4}) holds for $N+1$ and thus concludes the proof.
\end{IEEEproof}


\end{document}